\documentclass[preprint,twocolumn,3p]{elsarticle}
\usepackage{lineno,hyperref}
\modulolinenumbers[5]
\journal{Nuclear Instruments and Methods in Physics Research A}
\newcommand{\gx}{\textsc{GlueX}}
\usepackage{amsmath}
\usepackage{float}
\usepackage{caption}
\usepackage{booktabs}
\usepackage{subfigure}

\begin{document}

\begin{frontmatter}
\title{The \gx{} Start Counter Detector}	
\author[jlab,fiu]{E. Pooser\corref{correspondingauthor}}
\cortext[correspondingauthor]{Corresponding Author}
\ead{pooser@jlab.org}
\author[jlab]{F. Barbosa}
\author[fiu]{W. Boeglin}
\author[jlab]{C. Hutton}
\author[jlab]{M.M. Ito}
\author[fiu]{M. Kamel}
\author[fiu]{P. Khetarpal}
\author[fiu]{A. LLodra}
\author[jlab]{N. Sandoval}
\author[jlab]{S. Taylor}
\author[jlab]{T. Whitlatch}
\author[jlab]{S. Worthington}
\author[fiu]{C. Yero}
\author[jlab]{B. Zihlmann}
\address[jlab]{Thomas Jefferson National Accelerator Facility, Newport News, VA 23606, USA}
\address[fiu]{Florida International University, Miami, FL, 33199, USA}
\begin{abstract}
The design, simulation, fabrication, calibration, and performance of the \gx{} 
Start Counter detector is described.  The Start Counter was designed to operate 
at integrated rates of up to 9~MHz with a timing resolution in the range of 
500 to 825~ps (FWHM).  The Start Counter provides excellent solid angle 
coverage, a high degree of segmentation for background rejection, and can be 
utilized in the level 1 trigger for the experiment.  It consists of a 
cylindrical array of 30 thin scintillators with pointed ends that bend towards 
the beam line at the downstream end. Magnetic field insensitive silicon 
photomultiplier detectors were used as the light sensors.
\end{abstract}
\end{frontmatter}

\begingroup
\let\clearpage\relax
\section{Introduction}

The \gx{} experiment, staged in Hall D at the Thomas Jefferson National 
Accelerator Facility (TJNAF), primarily aims to study the spectrum of 
photo-produced mesons with unprecedented statistics in search for gluonic 
degrees of freedom.  The coherent bremsstrahlung technique is implemented to 
produce a linearly polarized photon beam that impinges on a liquid 
$\mathrm{H_{2}}$ target. A Start Counter detector was fabricated to properly 
identify the photon beam buckets and to provide accurate timing information.
\section{Design} \label{sec:design}

In this section we discuss the details of the \gx{} Start Counter design 
including the scintillators, support structure, light sensors and read out 
electronics.

\subsection{Overview} \label{sec:design_overview}
The Start Counter detector (ST), shown in Fig.~\ref{fig:sttargetiso}, surrounds 
a 30~cm long liquid $\mathrm{H_{2}}$ target while providing $\sim$$90 \%\ 
\mathrm{of\ 4 \pi}$ solid angle coverage relative to the target center.
	\begin{figure}[!htb]
		\centering
		\includegraphics[width=1.0\columnwidth]{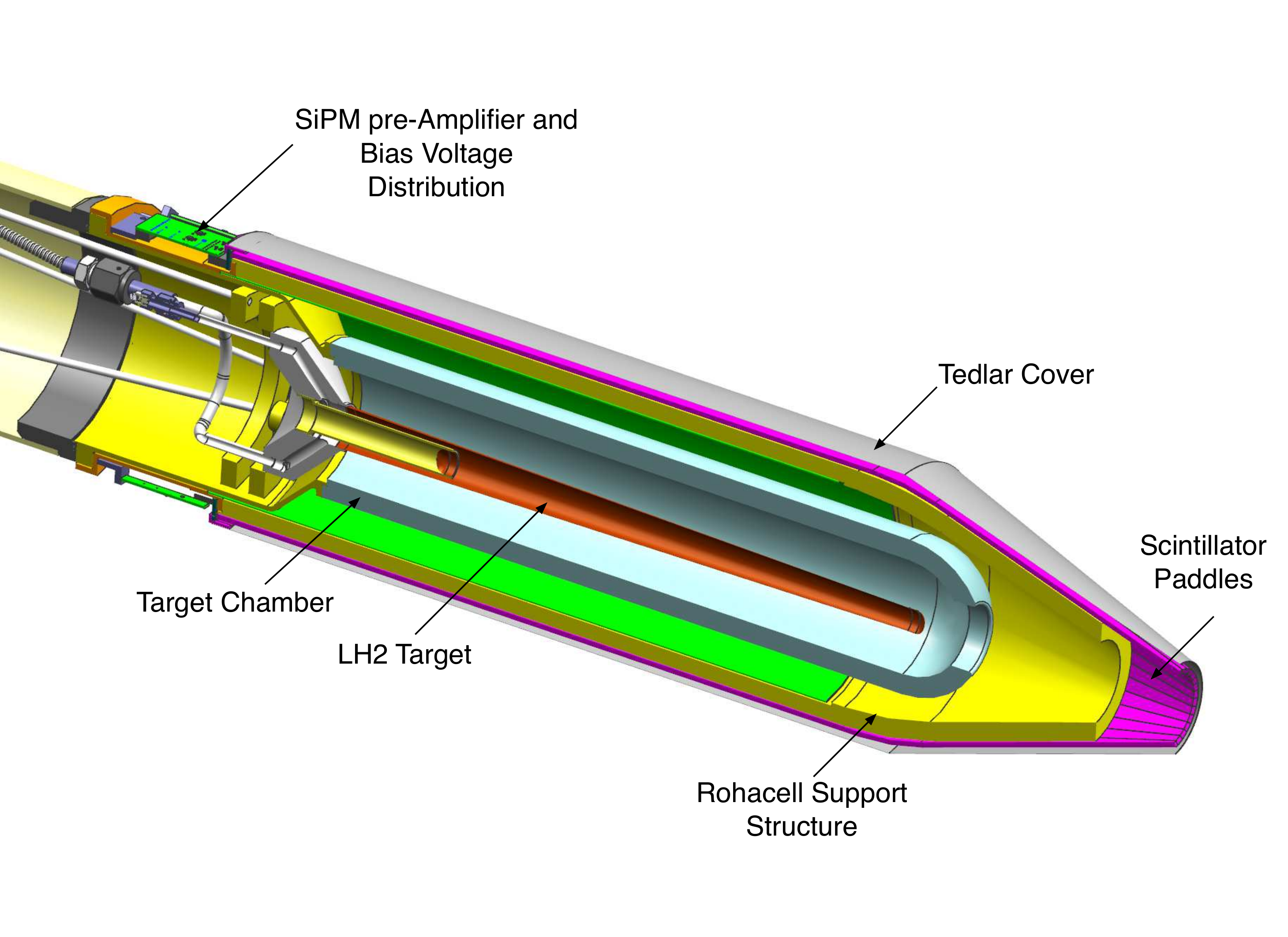}
		\caption{The \gx{} Start Counter mounted to the liquid $\mathrm{H_2}$ 
		target assembly.  The beam direction is oriented from left to right 
		down the central axis of the ST.}
		\label{fig:sttargetiso}
	\end{figure}
The primary purpose of the ST is to identify the photon beam bucket associated 
with the event reaction.  It is designed to operate at tagged photon beam 
intensities of up to $10^{8}\,\mathrm{\gamma/s}$ in the coherent peak where the 
photons range in energy from 8.4 to 9.0~GeV\cite{meyer_gluex_first_results}.  
The ST has a high degree of segmentation to limit the per paddle rates while 
also  providing background rejection information.  In order to resolve the 4~ns 
electron beam bunch structure delivered by the CEBAF to Hall-D with $6\sigma$ 
accuracy, the \gx{} Start Counter time resolution is required to be $<350\ 
\mathrm{ps}$.  It also facilitates particle  identification and can be utilized 
in the level 1 trigger of the \gx{} experiment during high luminosity  
running\cite{pooser16}\cite{somov_trig_talk}.

The ST has a cylindrical shape consisting of an array of 30 scintillators.  
Their pointed ends bend towards the beam line at the downstream end 
(Fig.~\ref{fig:st2dlabels}).
	\begin{figure}[!htb]
		\centering
		\includegraphics[width=1.0\columnwidth]{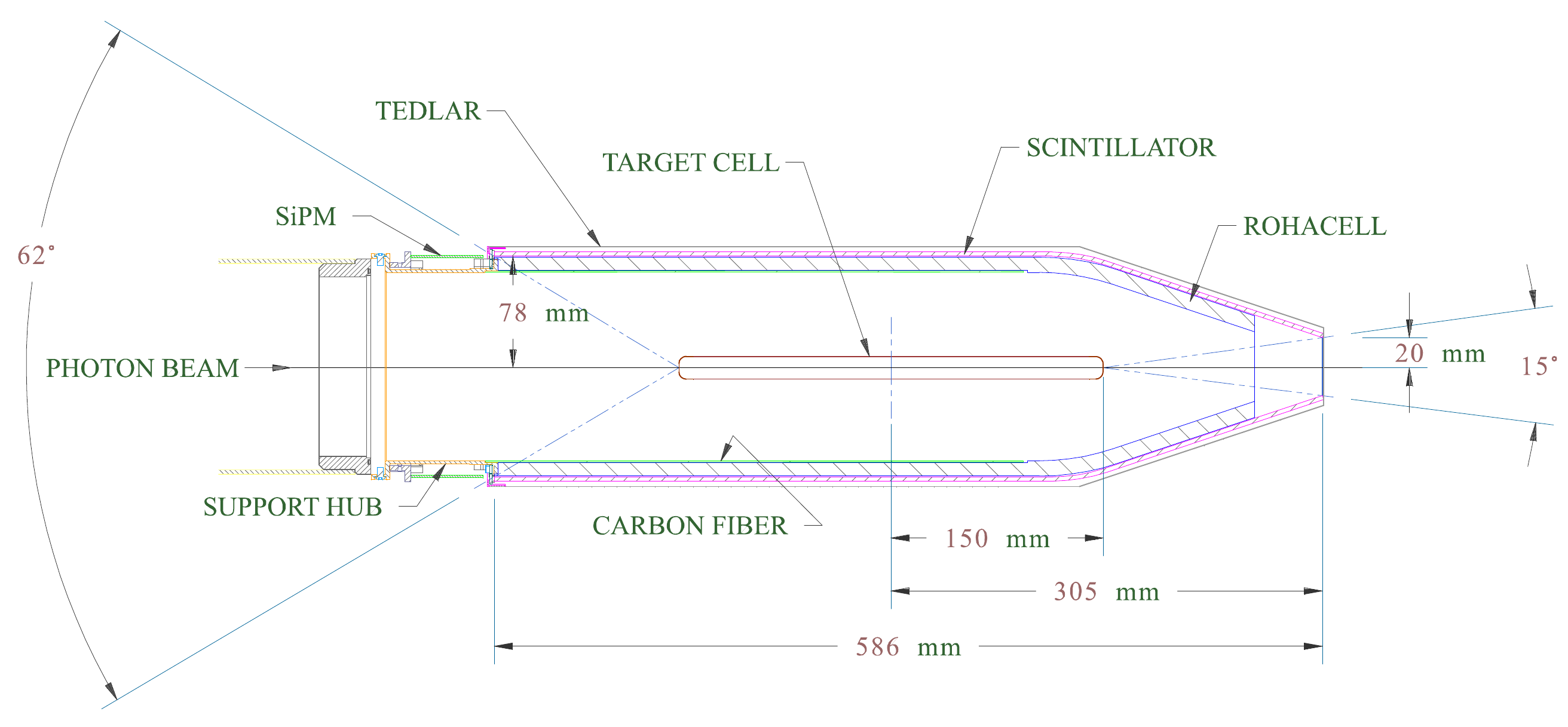}
		\caption{2-D cross section of the Start Counter.}
		\label{fig:st2dlabels}
	\end{figure}
EJ-200 scintillator material from Eljen Technology\cite{eljen} was 
selected for this application.  EJ-200 has a decay time of 2.1~ns and a long 
attenuation length\cite{ej200_specs}.  Silicon photomultiplier (SiPM) detectors 
were selected as the light sensors. These sensors are not affected by the 2~T  
magnetic field produced by the \gx{} superconducting solenoid magnet. The SiPMs 
were placed as close as possible to the upstream end of each scintillator 
element thereby minimizing the loss of scintillation light\cite{pooser16}.

\subsection{Scintillator Paddles} \label{sec:design_paddles}

Individual paddles were machined from long, thin, scintillator bars. Each  
paddle was manufactured to be 3~mm thick and diamond milled to be 600~mm in 
length and $\mathrm{20 \pm 2\ mm}$ wide.  The paddles were bent around a highly 
polished aluminum drum by applying localized infrared heating to the bend 
region.  The bent scintillator bars were then sent to McNeal Enterprises 
Inc.\cite{mcneal}, a plastic fabrication company, where they were machined to 
the desired geometry illustrated in Fig.~\ref{fig:stpaddleiso}.
	\begin{figure}[!htb]
		\centering
		\includegraphics[width=1.0\columnwidth]{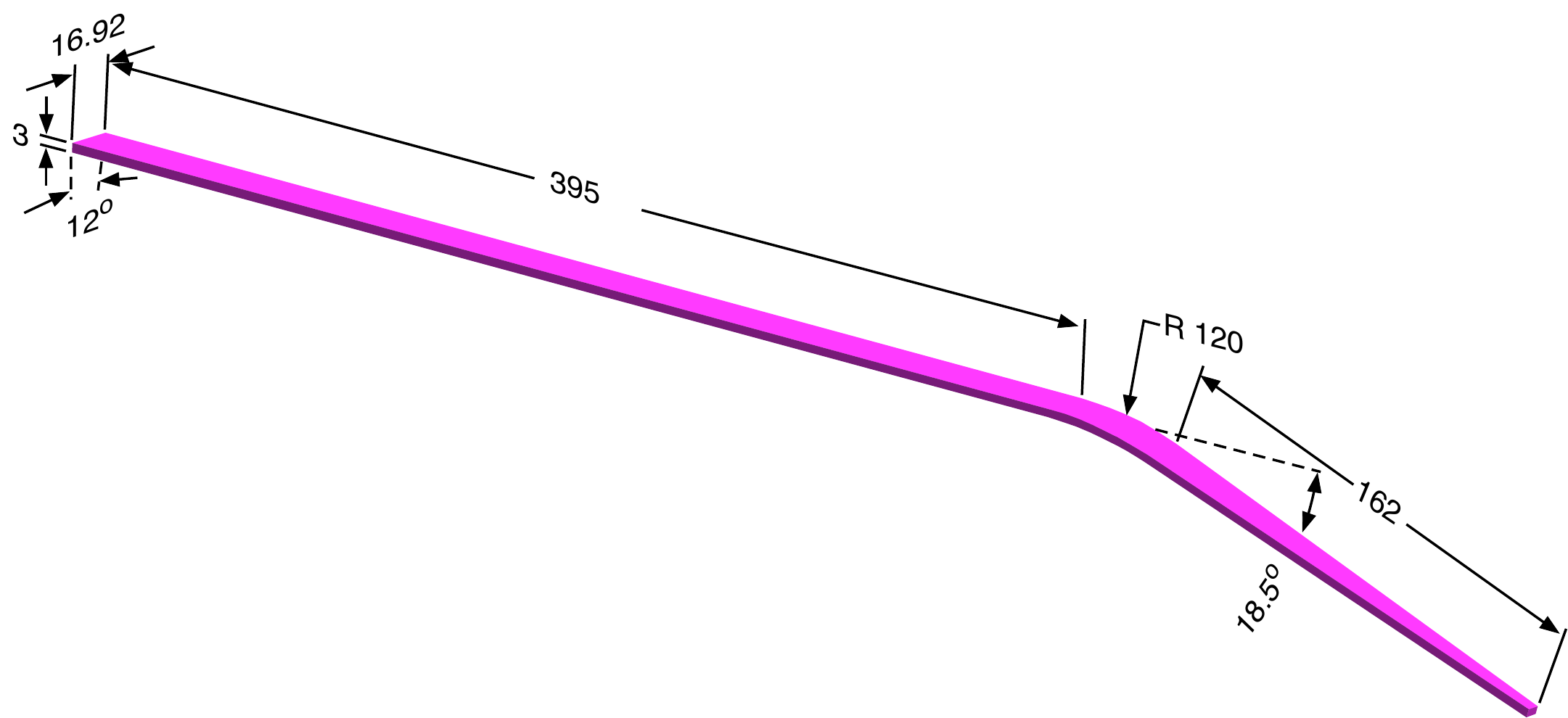}
		\caption{Start Counter single paddle geometry. Unlabeled dimensions shown are in mm.}
		\label{fig:stpaddleiso}
	\end{figure}

The paddles can be classified into three sections from the upstream to 
downstream end of the scintillator.  The straight section is 39.5~cm in length 
and is oriented parallel to both the target cell and beam line.  The bend 
region is a $18.5^{\circ}$ arc of radius 120~cm following the straight section. 
The tapered nose region is located past the target chamber and bends towards 
the beam line such that the tip of the nose is at a radial distance of 2~cm 
from the beam line.

After the straight scintillator bar was bent to the desired geometry the two 
flat surfaces, oriented orthogonal to the wide top and bottom surfaces, were 
cut at a $6^{\circ}$ angle.  During this process, the width of the top and 
bottom surfaces of the straight section were machined to be 16.92~mm and 
16.29~mm wide respectively. Thus, each of the paddles may be rotated 
$12^{\circ}$ with respect to the adjacent paddles so that they form a 
cylindrical shape with a conical end.  This geometrical design for the ST 
increases solid angle coverage while minimizing multiple scattering. 
 
\subsection{Support Structure} \label{sec:design_support}

The ST scintillator paddles are placed atop a low density 
Rohacell\cite{rohacell} foam support structure which envelopes the target 
chamber, illustrated in Fig.~\ref{fig:sttargetiso}.  The Rohacell is 11~mm 
thick and is rigidly attached to the support hub at the upstream end and 
extends along the length of paddles, partly covering the conical nose section.  
The cylindrical part of the Rohacell is reinforced with three layers of carbon 
fiber, each with a thickness of $\mathrm{650\ \mu m}$; this is illustrated in 
green in Figures~\ref{fig:st2dlabels} and \ref{fig:stmaterials}. 

The various layers of material that comprise the ST are illustrated in 
Fig.~\ref{fig:stmaterials}.
	\begin{figure}[!htb]
		\centering
		\includegraphics[width=1.0\columnwidth]{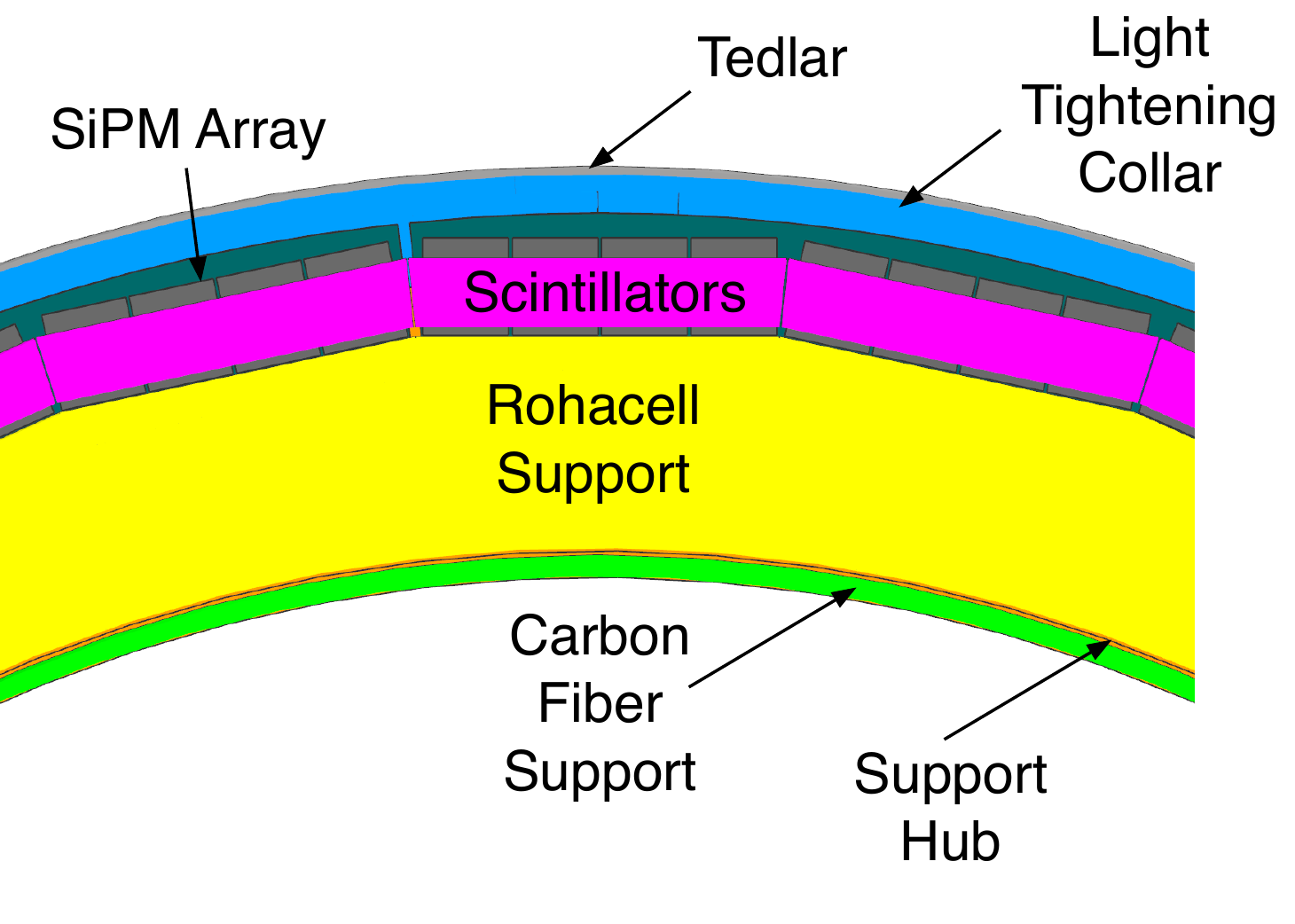}
		\caption{Start Counter materials.}
		\label{fig:stmaterials}
	\end{figure}
To ensure that the detector is light-tight, a plastic collar was placed around 
the top of the SiPMs at the upstream end.  The collar serves as a lip to which 
a cylindrical sheet of light insulation film (Tedlar\cite{tedlar_pvf}) is 
attached.  The nose section is covered by a cone of Tedlar which connects to 
the cylindrical section.  An additional cone of Tedlar is taped to the nose of 
the Rohacell and attached to the top Tedlar cone layer in order to ensure 
light-tightness.  A summary of the materials utilized in the ST are presented 
in Table~\ref{tab:st_mat_tbl}.
	\begin{table*}[!htbp]
		\centering
		\begin{tabular}{@{} l *5c @{}}
			\toprule
			\multicolumn{1}{c}{\textbf{Item}} & \textbf{Brand name} & 
			\textbf{Material} & \textbf{Thickness (mm)} & \textbf{Density 
			($\mathrm{g/cm^3}$)}  \\ 
			\midrule
			\textbf{Carbon fiber support} & Carbon fiber & Carbon & 1.950 & 
			1.523 \\ 
			\textbf{Rohacell support} & Rohacell & Polymethacrylimide & 11.0 & 
			0.075 & \\
			\textbf{Radial shims} & Kapton & Polyimide (type HN) & 0.762 & 1.42 
			& \\
			\textbf{Reflective film} & Aluminum foil & Aluminum & 0.016 & 2.70 
			& \\
			\textbf{Scintillator} & EJ-200 & Polyvinyltoluene & 3.0 & 1.023 & \\
			\textbf{Bundling wrap} & Stretch film & Polyethelene & 0.101 & 
			0.917 & \\
			\textbf{Light insulation film} & Tedlar & Polyvinyl Fluoride & 
			0.050 & 1.50 & \\
			\bottomrule
		\end{tabular}
		\caption{Start Counter materials.}
		\label{tab:st_mat_tbl}
	\end{table*}

\subsection{SiPM Readout Detectors} \label{sec:design_sipms}

Each scintillator bar is read out with an array of four magnetic field 
insensitive Hamamatsu S10931-050P multi-pixel photon counters 
(MPPCs)\cite{hamamatsu}.  Studies of several photo-detectors were performed in 
the initial design phase of the ST\cite{barbosa_sipm}. Based on these studies, 
the S10931-050P model was selected. An individual $\mathrm{3 \times 3\ mm^2}$ 
MPPC, here referred to as a ``SiPM'', consists of 3600 individual $\mathrm{50 
\times 50\ \mu m^2}$ avalanche photo-diode (APD) pixel counters operating in 
Geiger mode. The signal output from each SiPM is the sum of the outputs from 
all 3600 APD pixels\cite{sipm_spec}. 

The SiPM detectors are housed in a ceramic case that is surface mounted to a 
custom-fabricated printed circuit board (PCB).  The PCB is rigidly attached to 
the lip of the upstream support hub.  The individual ST scintillators are 
coupled to the SiPM arrays via an 250~$\mu$m air gap.

\subsection{Readout Electronics} \label{sec:design_electronics}

There are three primary components of the ST detector readout 
system.  The first component ``ST1'', shown in Fig.~\ref{fig:st1_mounted}, 
collects scintillation light from three paddles independently and distributes 
the bias voltages for the SiPMs.
	\begin{figure}[!htb]
		\centering
		\includegraphics[width=1.0\columnwidth]{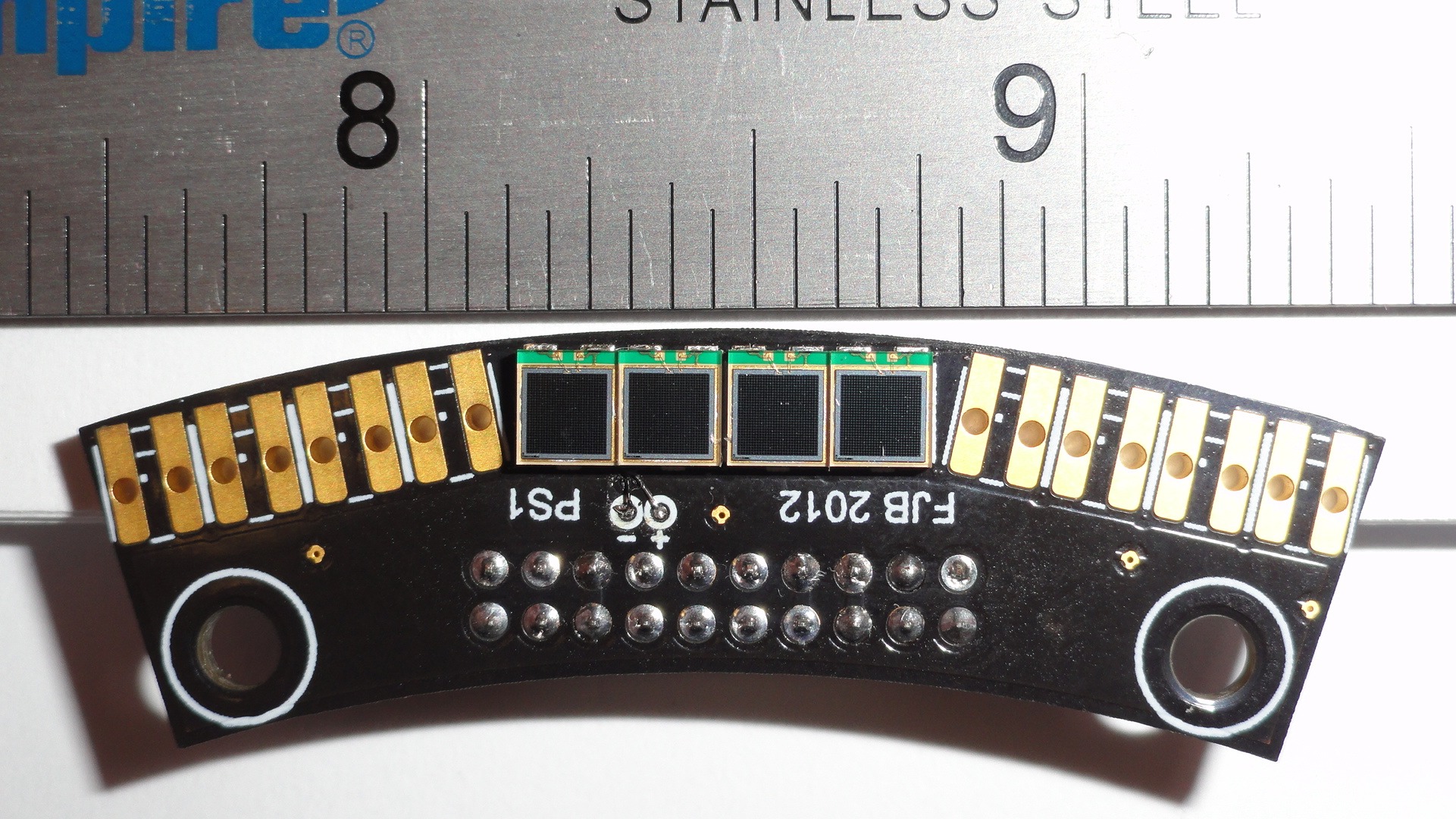}
		\caption{ST1 of Start Counter readout system.  Only the central array 
		is populated with SiPMs.  Approximately 72\% of the scintillator light 
		is collected at the upstream end.  The ST readout has 10 ST1 units in 
		total.  The ruler shown above is in inches.}
		\label{fig:st1_mounted}
	\end{figure}
Each array of four SiPMs has a thermocouple for temperature monitoring.

The second component ``ST2'', shown in Fig.~\ref{fig:stfullreadout}, has three 
pre-amplifiers, three buffers, and three factor-five amplifiers.  
	\begin{figure}[!htb]
		\centering
		\includegraphics[width=1.0\columnwidth]{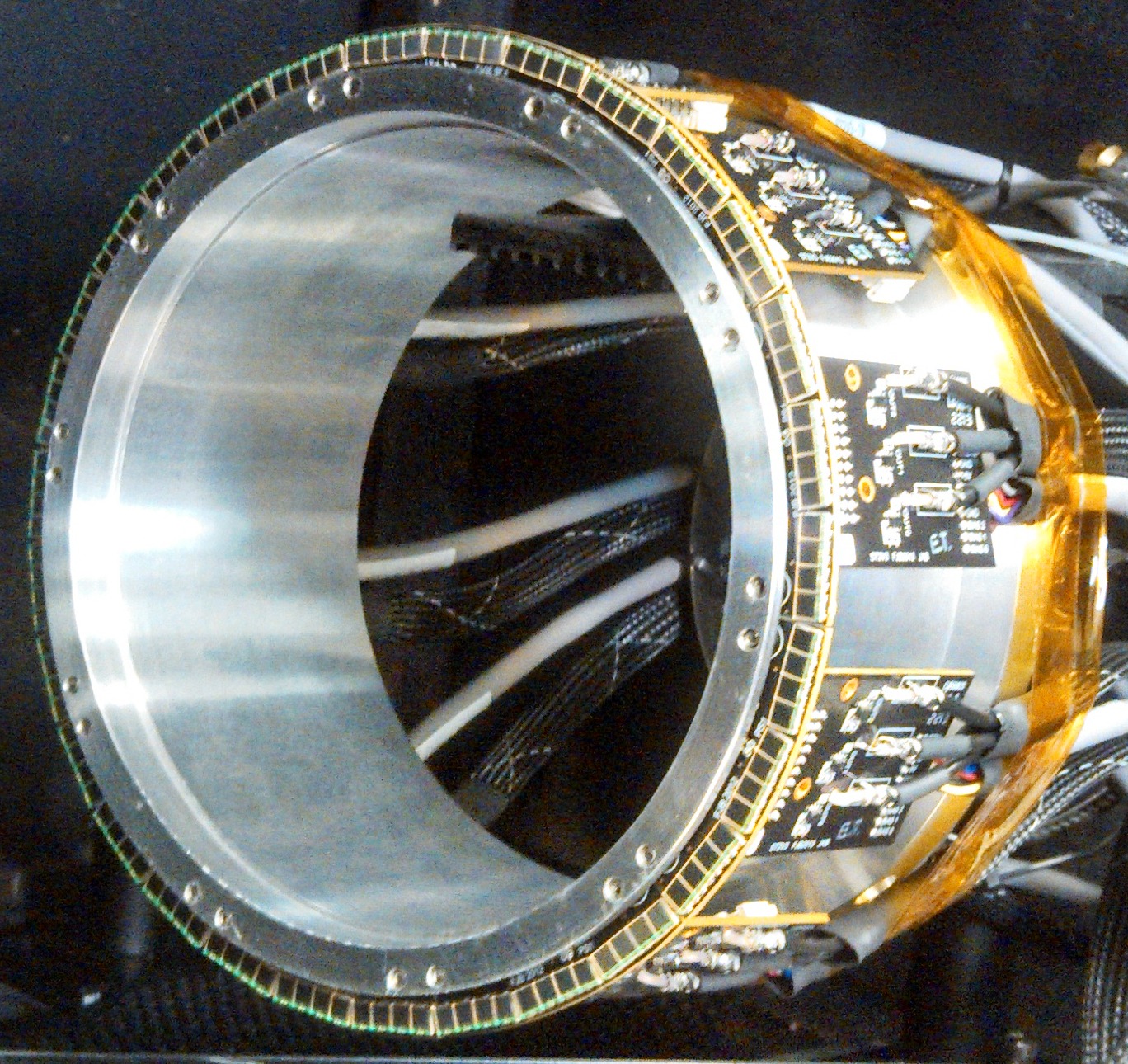}
		\caption{Fully assembled ST readout system.  The ST2 unit is connected 
		behind the ST1.  The full readout system is comprised of 10 ST2 units.}
		\label{fig:stfullreadout}
	\end{figure}
The output of each preamp is split; buffered for the analog-to-digital 
converter (ADC) output, and amplified for the time to digital converter (TDC) 
output.  The ADC outputs are digitized by Jefferson Lab VXS 250 MHz Flash ADC 
modules\cite{barbosa_fadc}.  The TDC outputs are input into Jefferson Lab 
leading edge discriminators, followed by a high resolution 32 channel Jefferson 
Lab VXS F1TDC V2 module\cite{barbosa_f1tdc}.  Furthermore, the ST2 has three 
bias distribution channels with individual temperature compensation via  
thermistors.  The ST2 is attached to the ST1 via a $90^{\circ}$ hermaphroditic 
connector.

The third component of the readout system, ``ST3'', provides the interface for 
the power and bias supplies.  It also routes the ADC and TDC outputs as well as 
the thermocouple output.  The ST3 is installed upstream of the Start Counter 
next to the beam pipe.  A schematic of the ST readout electronics is 
illustrated in Fig.~\ref{fig:Start Counter Electronics}.
	\begin{figure}[!htb]
		\centering
		\includegraphics[width=1.0\columnwidth]{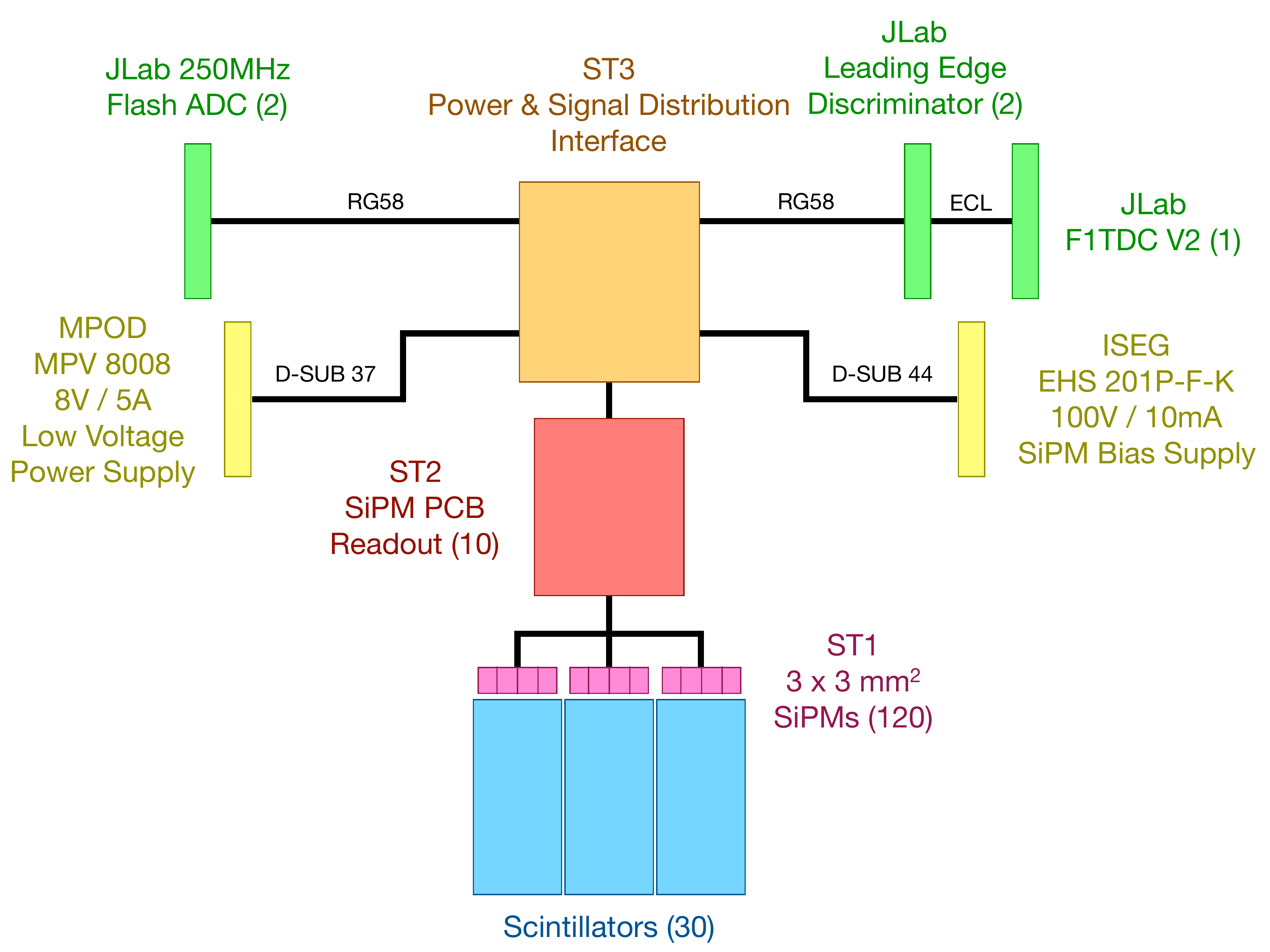}
		\caption{Start Counter readout electronics diagram.  Numbers in 
		parenthesis indicate the total for the system.}
		\label{fig:Start Counter Electronics}
	\end{figure}

\section{Simulation} \label{sec:sim}

In this section, Monte Carlo (MC) simulations of the performance and 
characteristics of machined scintillators are discussed.  These studies were 
performed using the Geant4 tool-kit, which simulates the passage of particles 
through matter \cite{geant4_website}.  Comparisons are made with data observed 
in experiments conducted on the bench (Sec.~\ref{sec:fab_test}) and with beam 
data (Sec.~\ref{sec:perform}).

\subsection{Simulating a Simplified Model of the ST} \label{sec:sim_simple}

As discussed in Sec.~\ref{sec:design_paddles}, the ST paddle geometry has a 
nose section which tapers at the downstream end. This causes the light 
collection efficiency of hits in the nose section to increase as the hit 
position moves farther from the photo-detectors, contrary to the usual behavior 
of scintillator material.

A simple Geant4 simulation was conducted to investigate the light collection 
efficiency. The details of the simulation are discussed in 
Ref.~\cite{pooser16}.  Only the two trapezoidal regions of a machined 
scintillator paddle were considered: the wide straight section and the 
tapered nose section. 

Ten thousand optical photons were generated at 16 different locations inside 
the medium of the scintillator. The photon energies ranged between 0.5 and 3.0 
eV\cite{krane_ch7} and were generated isotropically from points along a 3~mm 
path in the scintillator medium.  This path is oriented orthogonal to the wide 
surface of the scintillator.  The number of photons detected by the SiPM, 
denoted as ``SiPM Hits'', is shown in Figures~\ref{fig:sim_results} and 
\ref{fig:polished_vs_ground} as a function of the source locations.  For these 
studies, 100\% detection efficiency was assumed for the simulated SiPM.  In the 
case of the nose section, the SiPM was placed at the wider upstream end of the 
simulated scintillator bar.  The results for this simulation are presented in 
Fig.~\ref{fig:sim_results}.
	\begin{figure}[!htb]
		\centering
		\includegraphics[width=1.0\columnwidth]{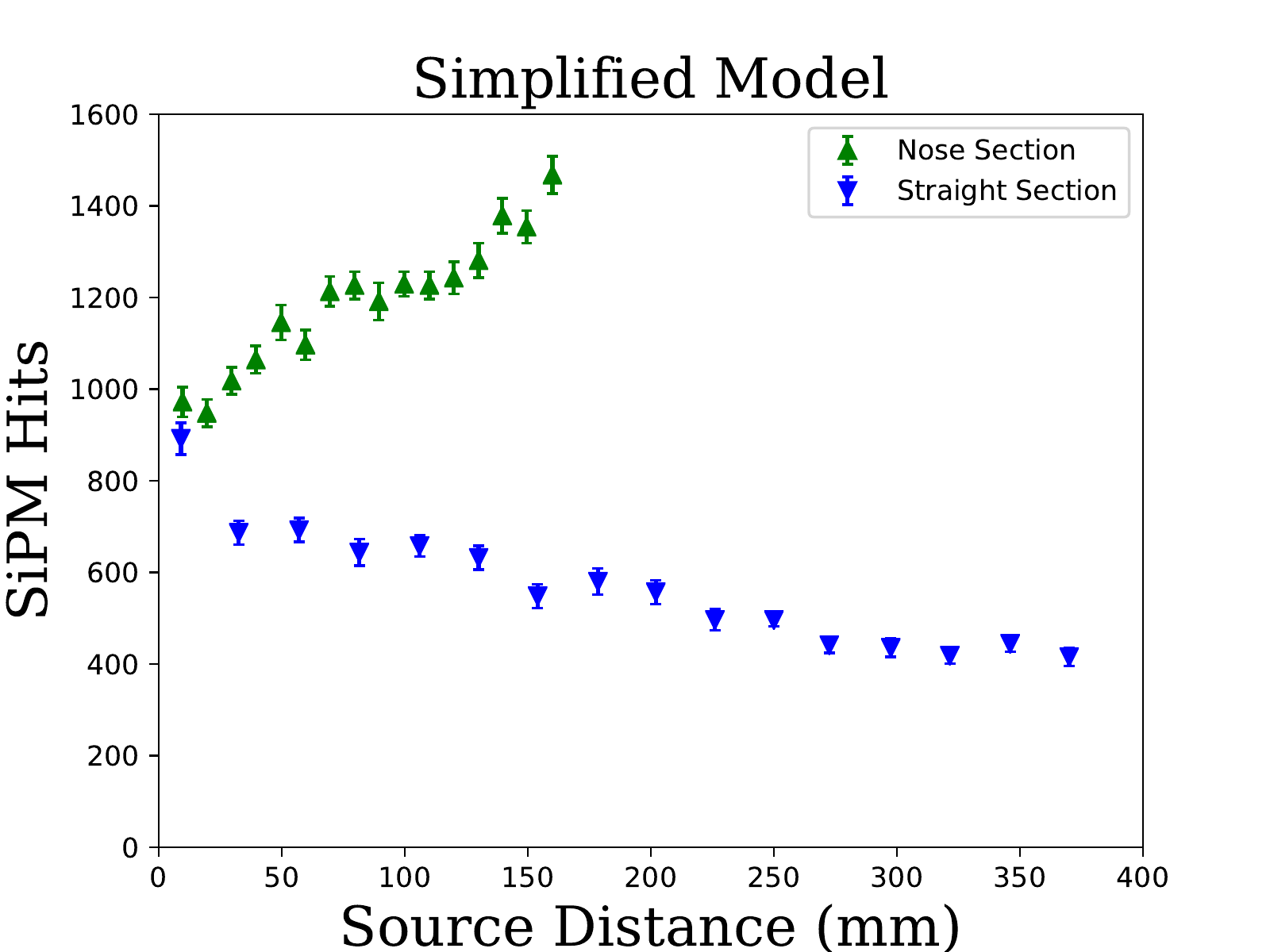}
		\caption{Simulation results for a simplified two section scenario. The 
		total number of photons which were collected by the SiPM detector for 
		each of the 16 source locations is plotted against the source distance 
		from the photon detector.}
		\label{fig:sim_results}
	\end{figure}

The simulation shows that the tapering trapezoidal geometry of the nose section 
results in improved light collection as the source moves further away from the 
readout detector.  There is an increase of 50\% in light collection as the 
source is moved from the near end to the far end of the nose section.  The 
quasi-rectangular straight section shows the typical loss off light yield as 
the source moves away from the photon detector.

\subsection{Simulating Machined Scintillator Geometry} \label{sec:sim_mach}

Further simulations were conducted to study the effects of the ST scintillator 
geometry and optical surface quality on light collection.  The scintillator 
geometry was imported into Geant4 from a Vectorworks CAD drawing utilizing the 
CADMesh utility\cite{cadmesh_g4}. The SiPM was modeled as a $12 \times 12 
\times 10\ \mathrm{mm^{3}}$ volume with a 100~$\mathrm{\mu m}$ air gap between 
it and the wide end of the straight section.  The volume surrounding the 
scintillator was defined to be air.  The scintillator material, SiPM photon 
detector, and optical photons were defined in an a manner identical to that 
discussed in Sec.~\ref{sec:sim_simple}.

To simulate the imperfections of scintillator surfaces, an optical surface 
``skin'' was defined.  The ``skin'' conformed to the POLISH and UNIFIED physics 
models\cite{scint_surface_sim} and was of the type ``dielectric-dielectric''.  
Both the transmission efficiency and reflection parameters were implemented as 
free parameters in order to study their various effects on light transmission. 

The POLISH model simulates a perfectly polished surface while the UNIFIED model 
defines the finish of the scintillator surface both of which are 
illustrated in Fig.~\ref{fig:polished_vs_ground} 
\cite{scint_surface_sim}. 
	\begin{figure}[!htb]
		\centering
		\includegraphics[width=1.0\columnwidth]{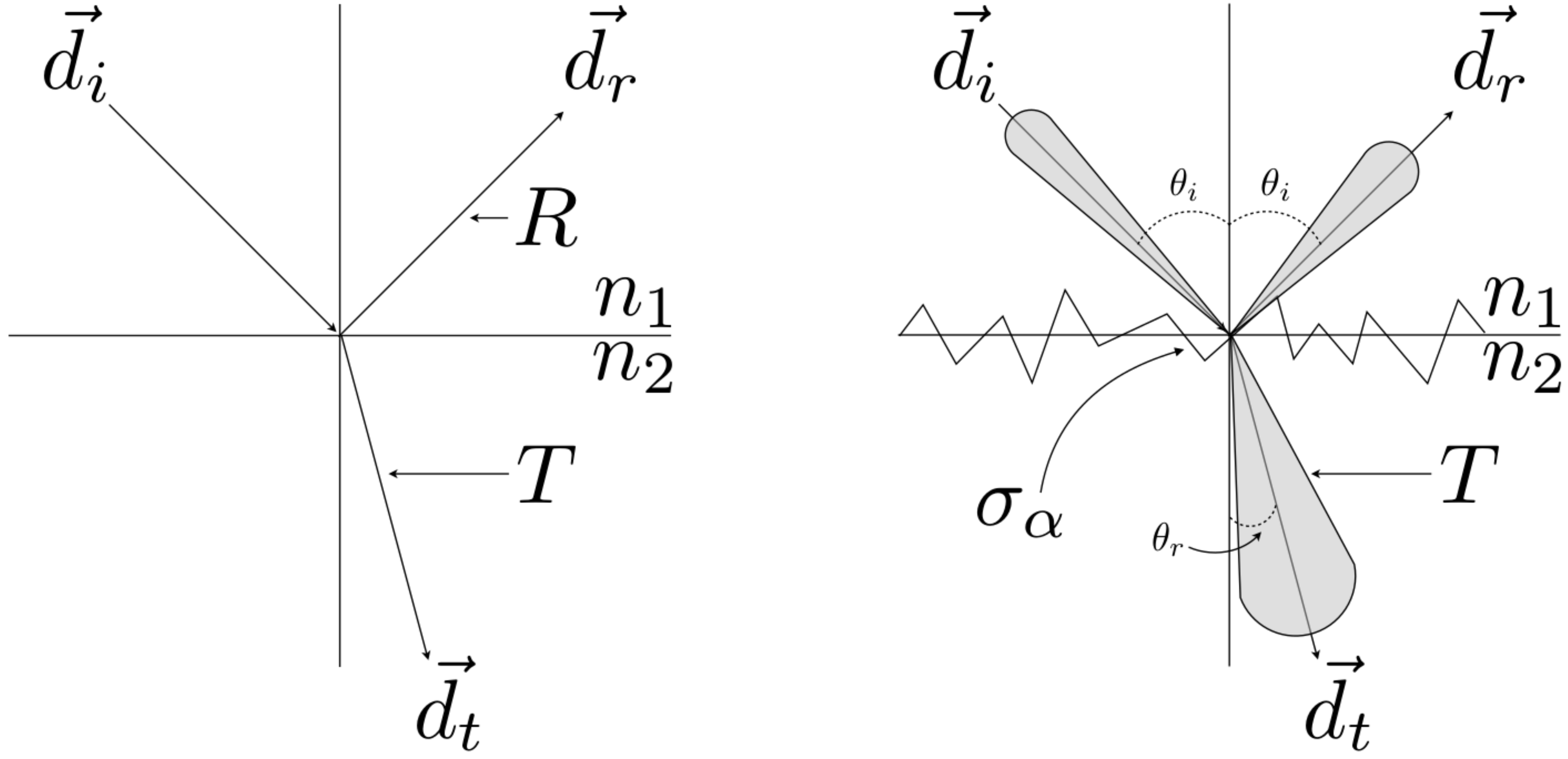}
		\caption{POLISH and UNIFIED models of scintillator surfaces.  Left: 
		Polar plot of the radiant intensity of the POLISH model.  Right: Polar 
		plot of	the radiant intensity in the UNIFIED model 
		\cite{scint_surface_sim}.  $\vec{d_i}, \vec{d_r}, \vec{d_t}$ are the 
		incident, reflected, and refracted photon direction vectors 
		respectively while $\vec{\sigma_i}$ and $\vec{\sigma_r}$ are the 
		associated incident and reflected angles with respect to the average 
		normals.  $n_1$ and $n_2$ are the indices of refraction for the 
		incident and transmission mediums respectively.  $R$ is the probability 
		of Fresnel reflection at the surface and the complementary probability 
		of transmission is simply $T = 1 - R$.}
		\label{fig:polished_vs_ground}
	\end{figure}
The details of the UNIFIED model parameters are discussed in detail in 
References \cite{pooser16} \& \cite{scint_surface_sim}.

As described in section \ref{sec:sim_simple}, 10,000 optical photons were 
generated in the scintillator medium every 2.5 cm and the number of hits in the 
SiPM were recorded.  For the POLISH model, only the transmission efficiency 
$\epsilon$ was varied.  For the UNIFIED model, $\epsilon$ and the radiant 
intensity parameters were held constant while $\sigma_{\alpha}$, which 
characterizes the standard deviation of the surfaces micro-facet orientation, 
was varied.  In both instances the attenuation length $\alpha$ was extracted in 
the straight section.  The results are shown in Fig.~
\ref{fig:transm_eff_vs_sig_alpha}. 
	\begin{figure*}[!htb]
		\centering
		\includegraphics[width=1.0\textwidth]{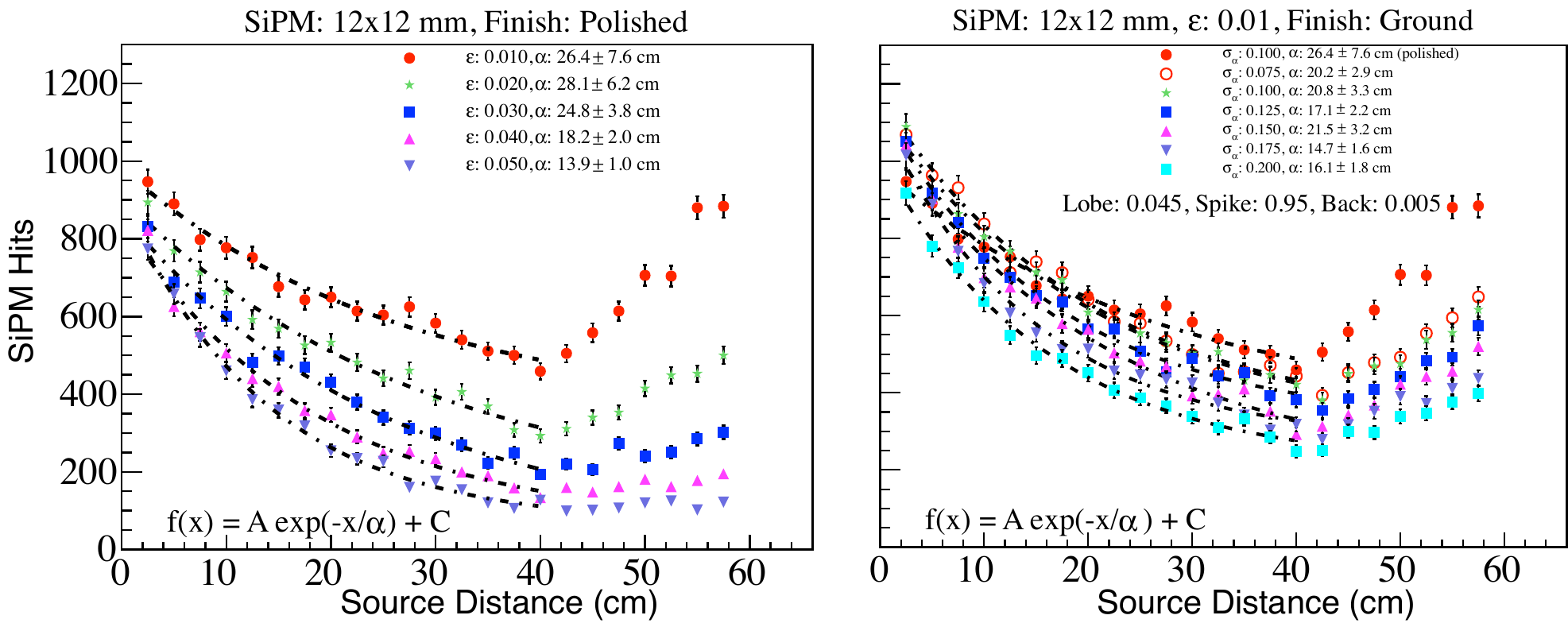}
		\caption{POLISH and UNIFIED model results.  Shown is the number of hits 
		recorded in the SiPM (vertical axis) versus the source distance 
		(x-axis).  Left: POLISH model varying the transmission efficiency 
		$\epsilon$.  Right: UNIFIED model varying the standard deviation of the 
		surfaces micro-facet orientation $\sigma_{\alpha}$.} 
		\label{fig:transm_eff_vs_sig_alpha}
	\end{figure*}

For the POLISH model it is clear that if the transmission efficiency increases, 
i.e. the reflection efficiency decreases, the amount of light collected in the 
SiPM decreases as illustrated in Fig.~\ref{fig:transm_eff_vs_sig_alpha}.  
Similarly, as the number of micro-facet orientations increase, meaning a more 
coarsely ground surface, the amount of light collection in the SiPM also 
decreases.  Moreover, good surface quality enhances the rise in light 
collection in the nose region.
\section{Misalignment Studies} \label{sec:misalign}

Here we discuss the relative alignment of a scintillator paddle with a SiPM 
detector and its effects on light collection and time resolution. 

\subsection{Experimental Set-up} \label{sec:misalign_setup}

The SiPM was mounted atop a Newport MT-XYZ (MT) linear translation 
stage\cite{newport_mt_xyz} with adjustment screws providing translations of 
$318\ \mathrm{\mu m}$ per $360^{\circ}$ rotation.  The SiPM collected light 
from a scintillator paddle at the upstream end of the straight section.  A 
$\mathrm{^{90}Sr}$ source and trigger photomultiplier tube (PMT) were fixed 
$\mathrm{24.5~cm}$ downstream from the readout end.  The response of the SiPM 
was recorded as a function of its relative position to the scintillator.  

Utilizing a camera, the vertical and horizontal alignment of the SiPM relative 
to the scintillator was measured with $25~\mathrm{\mu m}$ accuracy.
	\begin{figure}[!htb]
		\centering
		\includegraphics[width=1.0\columnwidth]{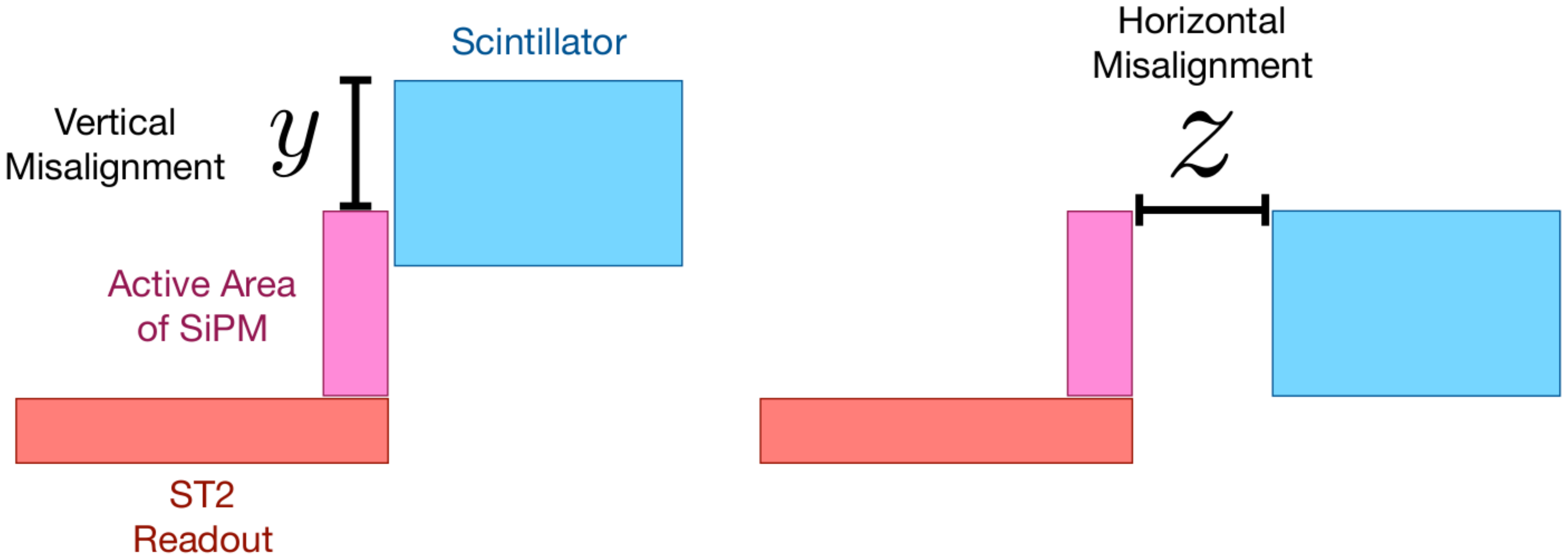}
		\caption{Optics setup for misalignment studies.  Left: SiPM \& 
		scintillator vertical misalignment. Right: SiPM \& scintillator 
		horizontal misalignment.}
		\label{fig:sipm_va_optics} 
	\end{figure}
Further details of the experimental set-up are discussed in 
Ref.~\cite{pooser16}. 

\subsection{Vertical Alignment of SiPM and Scintillator}
\label{sec:misalign_vert}

The scintillator remained fixed while the SiPM was scanned across the upstream 
end of the scintillator (Fig.~\ref{fig:sipm_va_optics}). During this scan, the 
horizontal alignment ($z$) of the SiPM and scintillator was fixed at a distance 
of $\mathrm{100\ \mu m}$ and was monitored closely.  At $y = 0$ the SiPM and 
scintillator are aligned vertically.  The measurements and simulations are 
shown in Fig.~\ref{fig:sipm_va}. 
	\begin{figure}[!htb]
		\centering
		\includegraphics[width=1.0\columnwidth]{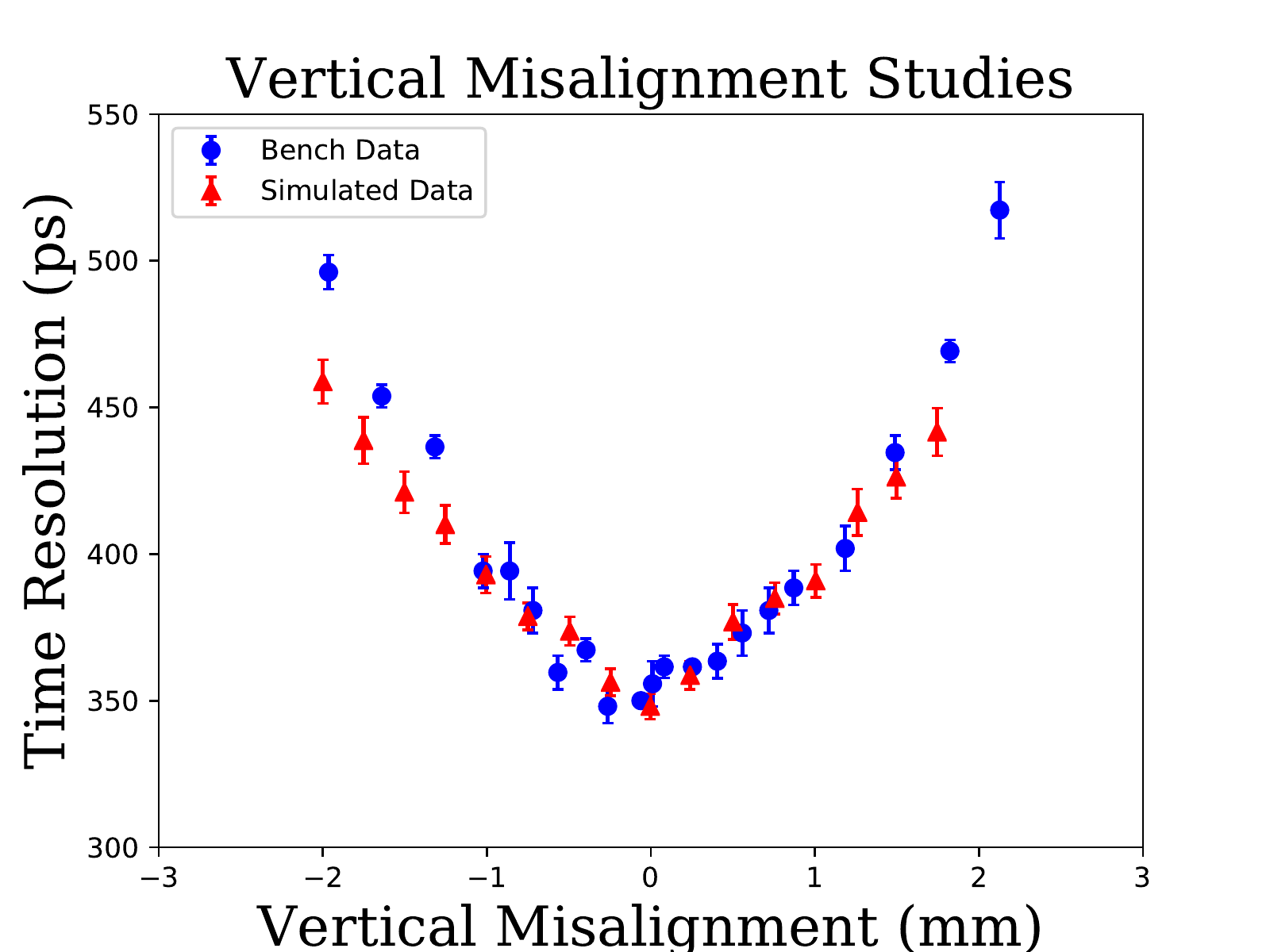}
		\caption{Vertical misalignment results.  The minimum time resolution 
		obtained was approximately 350 ps which was expected.  Once the SiPM 
		exceeded $y = \pm 3\ mm$, no active area of the SiPM was directly 
		coupled 
		to the face of the scintillator.}  
		\label{fig:sipm_va} 
	\end{figure} 
There is no significant variation of time resolution within a $\pm 300\ 
\mathrm{\mu m}$ range of the optimal alignment. 

A Geant4 simulation, done in a manner similar to that discussed in section 
\ref{sec:sim_mach}, was utilized to study the effect of vertical misalignment.  
The photon collection statistics at various $y$-positions in simulation matched 
data taken on the bench.  Ergo, the measured time resolution is dominated by 
photon collection statistics.  Thus, we determined the simulated time 
resolutions empirically, by scaling light collection to the time resolutions 
measured on the bench.  The acceptable range of vertical misalignment is 
approximately $\pm 250\ \mathrm{\mu m}$.

\subsection{Horizontal Alignment of SiPM and Scintillator}

The effects of varying the horizontal alignment were also studied.  While the 
horizontal alignment ($z$) was varied, the vertical alignment ($y$) was kept 
constant at the optimal location ($y = 0$), and was monitored both 
optically and manually with a micrometer. 

The SiPM was moved along the $z$-axis.  We defined $z = 0$ to be the position 
where the active area of the SiPM was flush against the face of the 
scintillator paddle.  The results of this study are illustrated in Fig. 
\ref{fig:sipm_coupling}. 
	\begin{figure}[!htb]
		\centering
		\includegraphics[width=1.0\columnwidth]{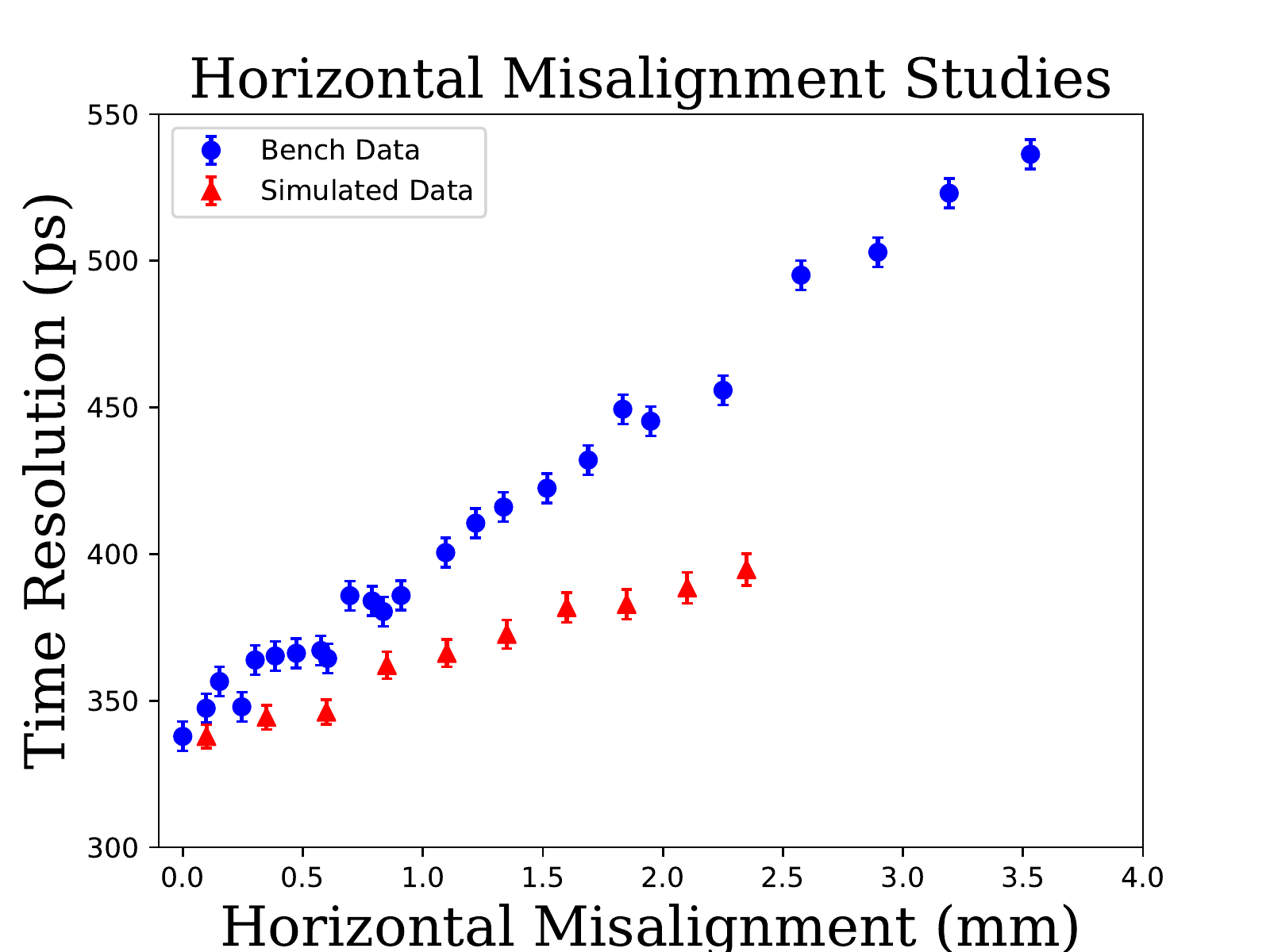}
		\caption{Horizontal misalignment results.}
		\label{fig:sipm_coupling}
	\end{figure}
While the simulation underestimates the degradation of resolution with 
increasing horizontal alignment, it is clear from the data that the optimal 
coupling range is $z < 350\ \mu \mathrm{m}$.  Moreover, there is no 
significant 
degradation in time resolution for $z < 600\ \mu \mathrm{m}$.
\section{Fabrication} \label{sec:fab}

The details of polishing and characterizing machined scintillators, as well as 
the construction of the Start Counter are discussed.

\subsection{Polishing Machined Scintillators} \label{sec:fab_polish}

While undergoing edge polishing at McNeil Enterprises, the machined 
scintillators incurred surface damage and were exposed to chemical contaminants 
known to harm scintillator surfaces. Polishing was required to restore adequate 
performance characteristics.

To polish the machined scintillator surfaces, Buehler Micropolish II 
deagglomerated $\mathrm{0.3\ \mu m}$ alumina suspension was utilized 
\cite{buehler}.  The polishing suspension was diluted with a 5:1 ratio of 
de-ionized $\mathrm{H_{2}O}$ to alumina and applied to a cold, wet $6'' \times 
0.5''$ Caswell Canton flannel buffing wheel \cite{caswell} operated at 
speeds less than 1500~RPMs. The surfaces of the scintillators were carefully 
buffed until the large surface defects were removed. In order to eliminate 
small localized surface defects, hand polishing with a soft NOVUS premium 
Polish Mate microfilament cloth \cite{novus} and diluted polishing suspension 
was applied.  These polishing procedures made the scintillators void of 
most visible scratches and surface defects. 

The improved surface quality of the polished scintillators are shown in 
Fig.~\ref{fig:polshing_effects} where a scintillator paddle before and after 
polishing is shown. 
	\begin{figure}[!htb]
		\centering
		\includegraphics[width=1.0\columnwidth]{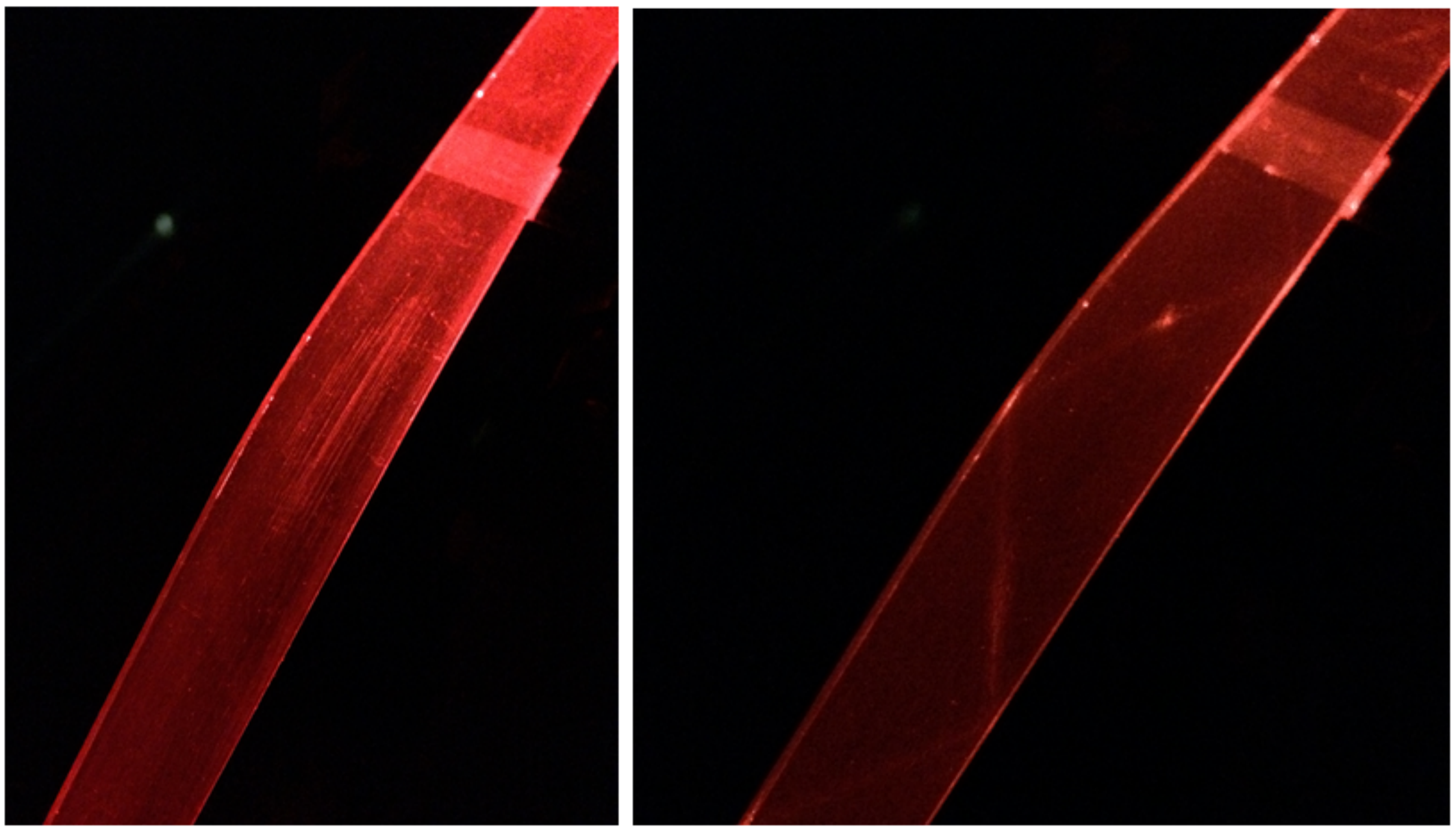}
		\caption{Effects of polishing scintillators. Left: non-diffuse laser 
		incident on an edge, before polishing, at the upstream end of the 
		straight section. Right: non-diffuse laser incident on the same edge, 
		after polishing, at the upstream end of the straight section.} 
		\label{fig:polshing_effects} 
	\end{figure}
A red laser beam was shone into the scintillator medium from the upstream end 
aimed at one edge.  The unpolished scintillator had such poor surface quality 
that the reflections of the laser in the bend region could not be resolved.  
However, the reflections in the polished scintillator can clearly be observed 
traversing the bend and nose region.  On average, at the tip of the nose, the 
scintillators exhibited a 15\% improvement in time resolution.  Moreover, 
variation in performance from counter to counter was substantially reduced.

\subsection{Testing} \label{sec:fab_test}

The polished scintillators were tested for light output and time resolution 
properties. 
	\begin{figure}[!htb]
		\centering
		\includegraphics[width=1.0\columnwidth]{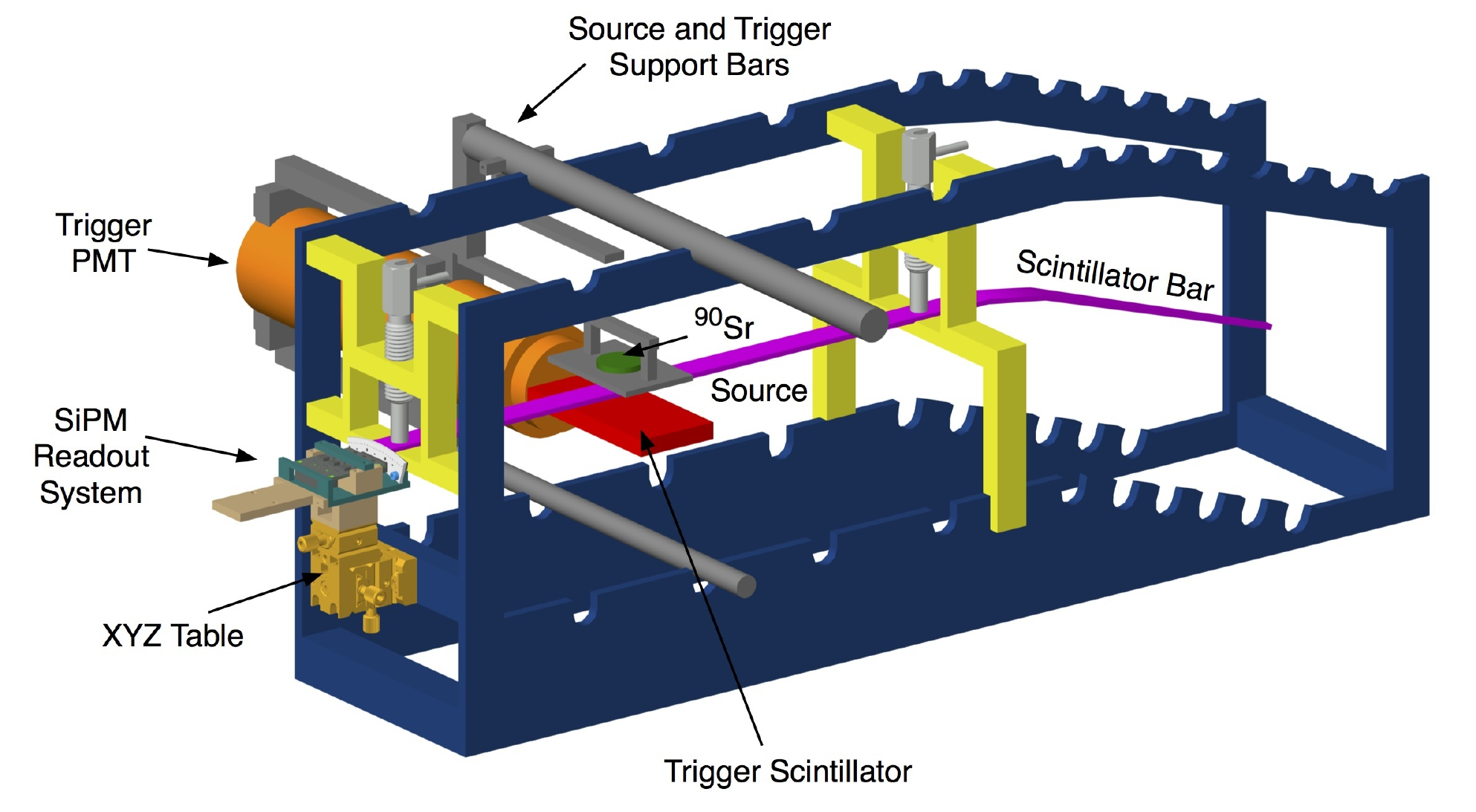}
		\caption{CAD Drawing of the scintillator test stand.}
		\label{fig:test_stand_model}
	\end{figure} 
A test stand (Fig.~\ref{fig:test_stand_model}) was used to measure the response 
of machined scintillators at four locations in the straight section, three in 
the bend, and five in the nose.   

The measurements were conducted with a collimated $\mathrm{^{90}Sr}$ source 
oriented orthogonal to the wide flat surface of the scintillators.  The 
$\mathrm{^{90}Sr}$ source provides electrons ranging from 
$\mathrm{0.5-2.3~MeV}$ in energy \cite{nndc_sr90}\cite{nndc_y90}.  A trigger 
PMT was placed underneath the scintillator on the opposite side and provided 
the TDC start and ADC gate.  A SiPM detector array identical the final ST 
assembly, collected light from the scintillator being tested.  The ADC and TDC 
data were analyzed to determine the light output and time resolution.

The 30 machined scintillator paddles that exhibited the best time resolution 
and light output properties from a set  of 50 were selected for the final 
construction.  These scintillators were then wrapped in $\mathrm{16.5~\mu m}$ 
thick reflective film (aluminum foil) and tested again.  Their measured time 
resolutions are illustrated in Fig.~\ref{fig:time_res_comp_final_30}.
	\begin{figure}[!htb]
		\centering
		\includegraphics[width=1.0\columnwidth]{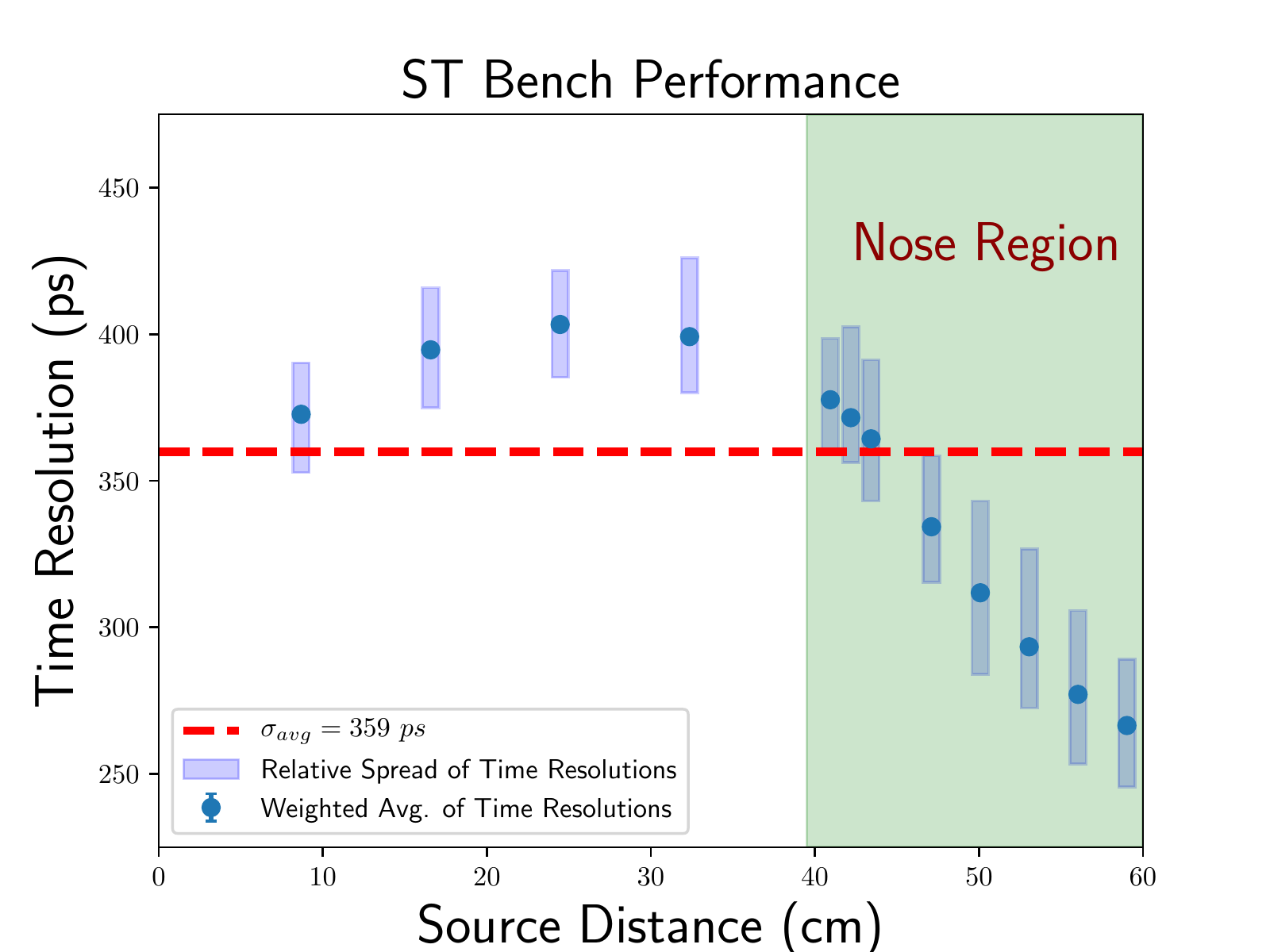}
		\caption{Weighted average of the time resolution of 30 scintillator 
		paddles as a function of distance from the SiPM.  The shaded 
		vertical blue boxes indicate the relative spread of the time 
		resolutions among the 30 paddles.  The dashed line indicates the 
		weighted average over the 12 data points.}
		\label{fig:time_res_comp_final_30}
	\end{figure}
The phenomenon of increased light collection in the nose region is observed.  
The larger time resolution in the straight section is due to light which 
initially travels downstream is reflected from the nose.

\subsection{Assembly} \label{sec:fab_ass}

To build the ST an assembly jig (Fig.~\ref{fig:ajcaddrawing}) was fabricated.
	\begin{figure}[!htb]
		\centering
		\includegraphics[width=1.0\columnwidth]{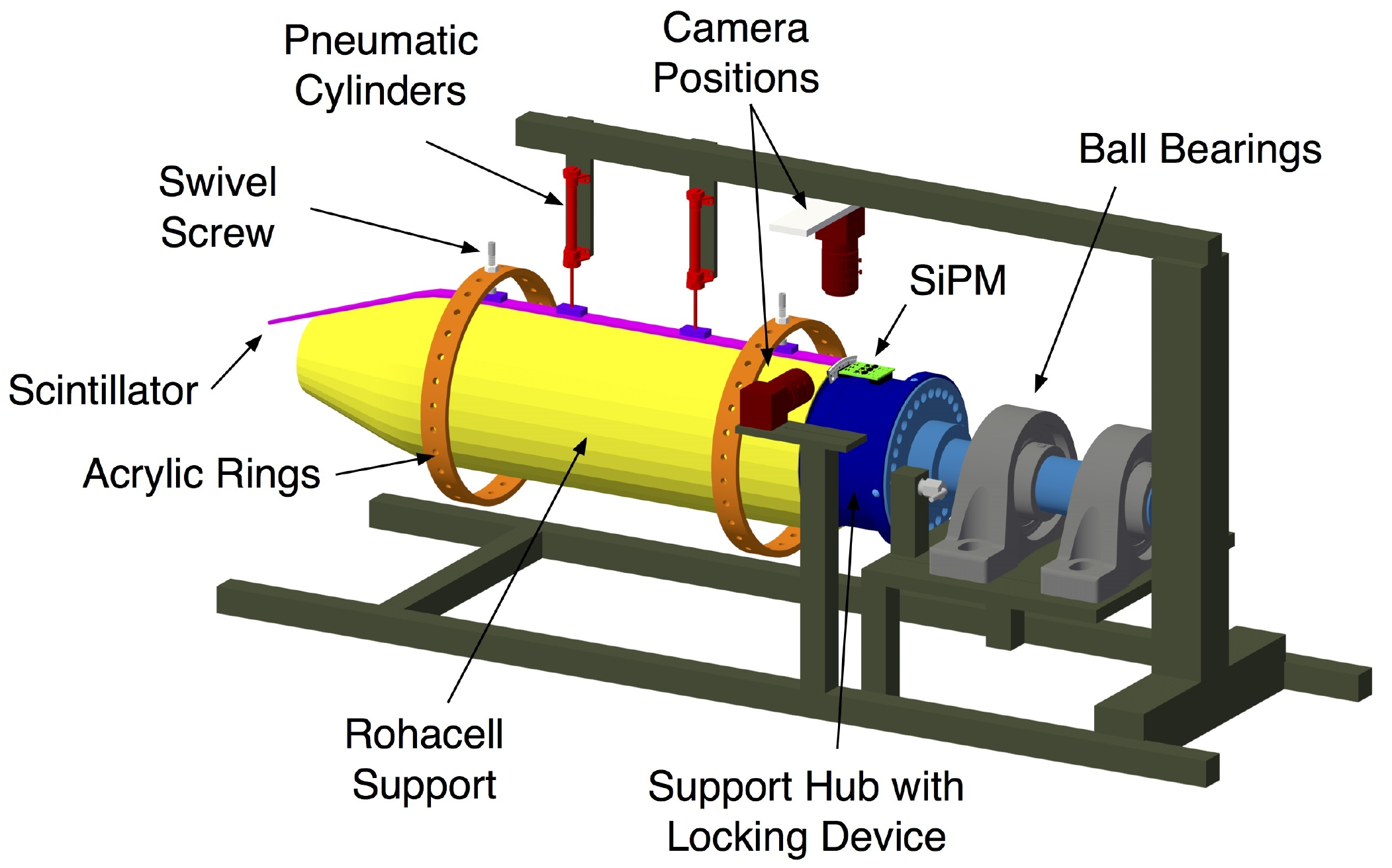}
		\caption{CAD drawing of the ST assembly jig.}
		\label{fig:ajcaddrawing}
	\end{figure} 
The upstream support hub and Rohacell support structure were attached to a 
rotating bracket that moved in discretized $12^{\circ}$ increments.  Two 
pneumatic cylinders with soft, semi-dense rubber feet were used to hold a 
single scintillator in place.  Two free floating acrylic rings, with 30 tapped 
holes $12^{\circ}$ apart, housed $10^{\circ}$ swivel pad thumb screws fitted 
with silicone foam.  The thumb screws held installed paddles in place.

A camera was used to measure and control the scintillator/SiPM vertical and 
horizontal alignments.  Vertical alignment was achieved by using Kapton shims 
between the scintillator and the support structure. The horizontal alignment 
was configured to a distance less than $\mathrm{200\ \mu m}$ between the 
scintillator and the SiPMs.

To secure paddles the to the Rohacell support structure the ST was wrapped 
around its circumference using self-adhesive transparent bundling wrap (0.8 mil 
thick, 6 in wide) at six locations along the length of the detector as seen in 
Fig.~\ref{fig:light_tightening_cone}. 
	\begin{figure}[!htb]
		\centering
		\includegraphics[width=1.0\columnwidth]{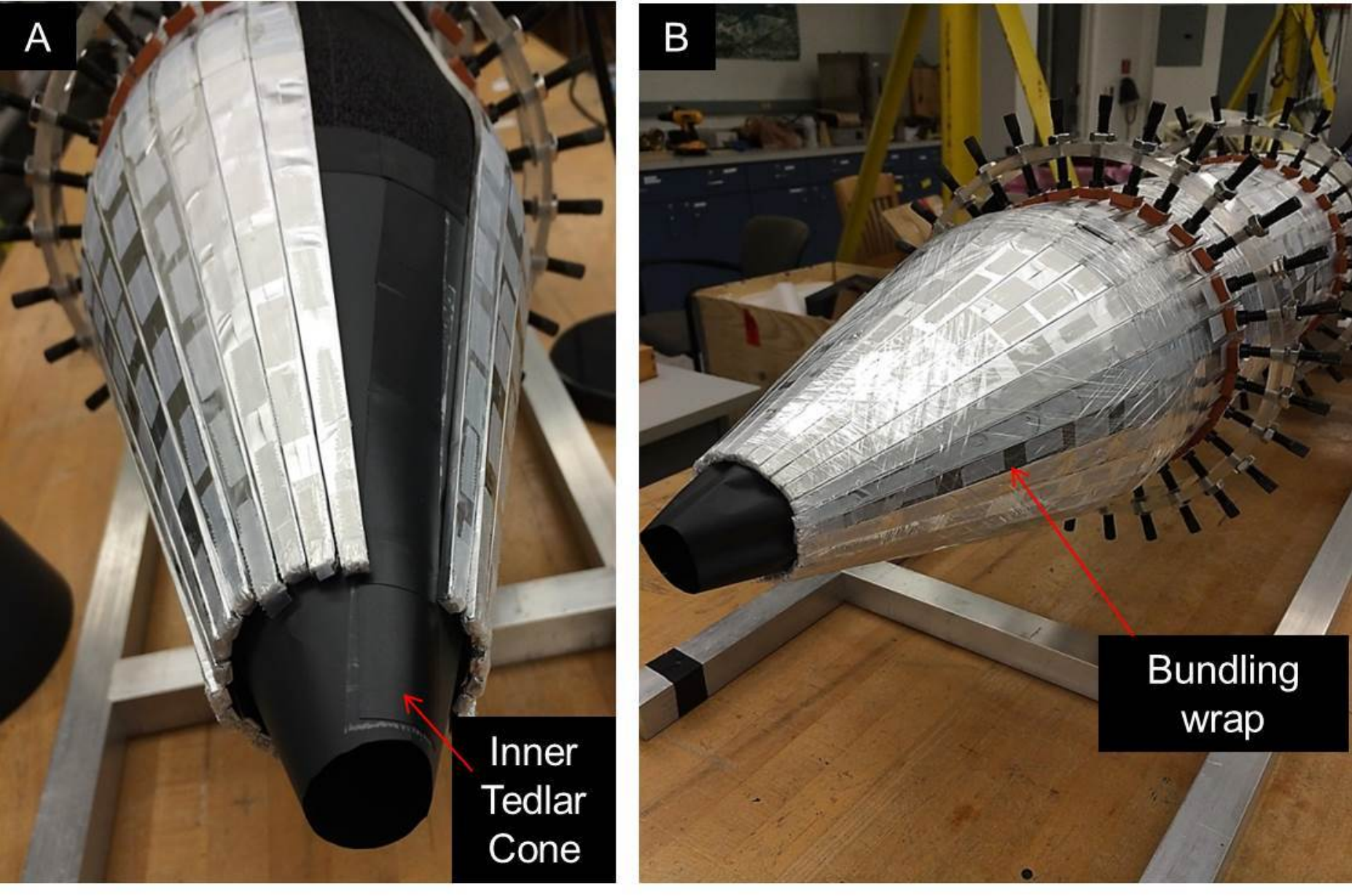}
		\caption{ST assembly before (A) and after (B) wrapping with bundling 
		wrap}
		\label{fig:light_tightening_cone}
	\end{figure}
Details regarding the assembly process are discussed in Ref.~\cite{pooser16}. 

The fully assembled ST is mounted around the \gx{} liquid 
$\mathrm{H_{2}}$ target as shown in Fig.~\ref{fig:light_tight_st}.
	\begin{figure}[!htb]
		\centering
		\includegraphics[width=1.0\columnwidth]{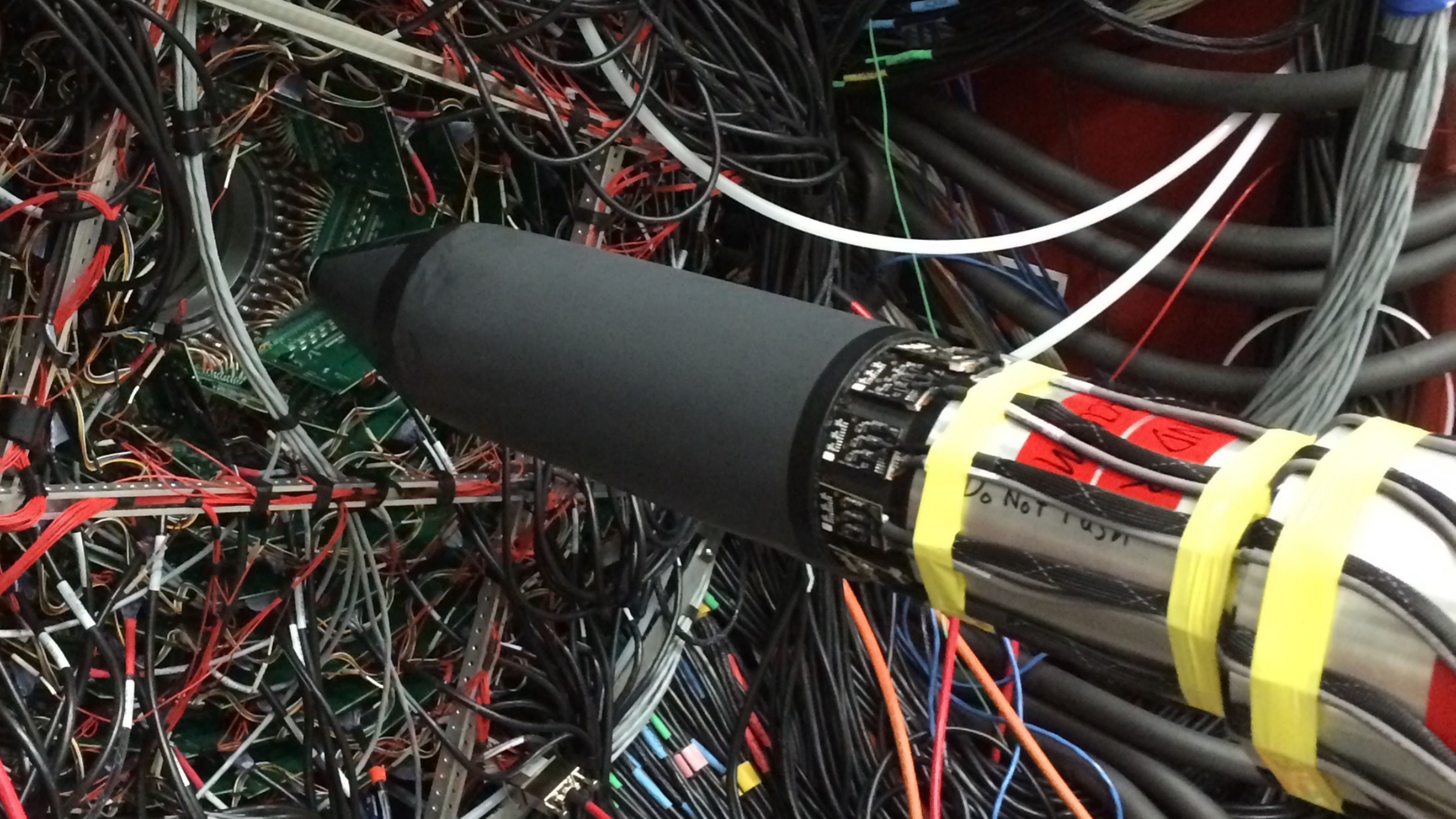}
		\caption{ST mounted the \gx{} target.  The beam direction 
		is from right to left and travels down the central axis of the ST.  
		During operation the ST resides in the bore of the central tracking 
		chamber, which is visible in the top left corner.}
		\label{fig:light_tight_st}
	\end{figure} 
\section{Calibration} \label{sec:calib}

The procedures to optimize the time resolution for particle identification 
(PID) and time of flight (TOF) are discussed here. 

\subsection{Time-walk Correction} \label{sec:calib_tw}

To correct the TDC timing for variations due to pulse shape, we use the timing 
signal from the FADC250’s. The latter uses a digital algorithm similar to a 
constant-fraction discriminator and therefore gives a time largely independent 
of pulse height\cite{pooser16}\cite{dong_fadc}.  The TDC/FADC time difference 
is given by Eq.~\ref{eq:tdc_adc_tdiff} where $i$ is the paddle number index.
	\begin{equation} \label{eq:tdc_adc_tdiff}
		\delta t_{i} = t^{\rm TDC}_{i} - t^{\rm FADC}_{i}
	\end{equation}
	
The FADC250's report the amplitude, integral, and time of the input analog 
signals\cite{dong_fadc}.  The amplitude was selected for the time-walk 
corrections because it is correlated better with the leading edge time of the 
pulse\cite{pooser16}.  Figure~\ref{fig:time_walk} (left) shows a typical 
time-walk spectrum, i.e. $\delta t$ versus the pulse amplitude, for 
one paddle of the ST. 
	\begin{figure*}[!htp]
		\centering
		\subfigure{\includegraphics[width=0.5\textwidth]{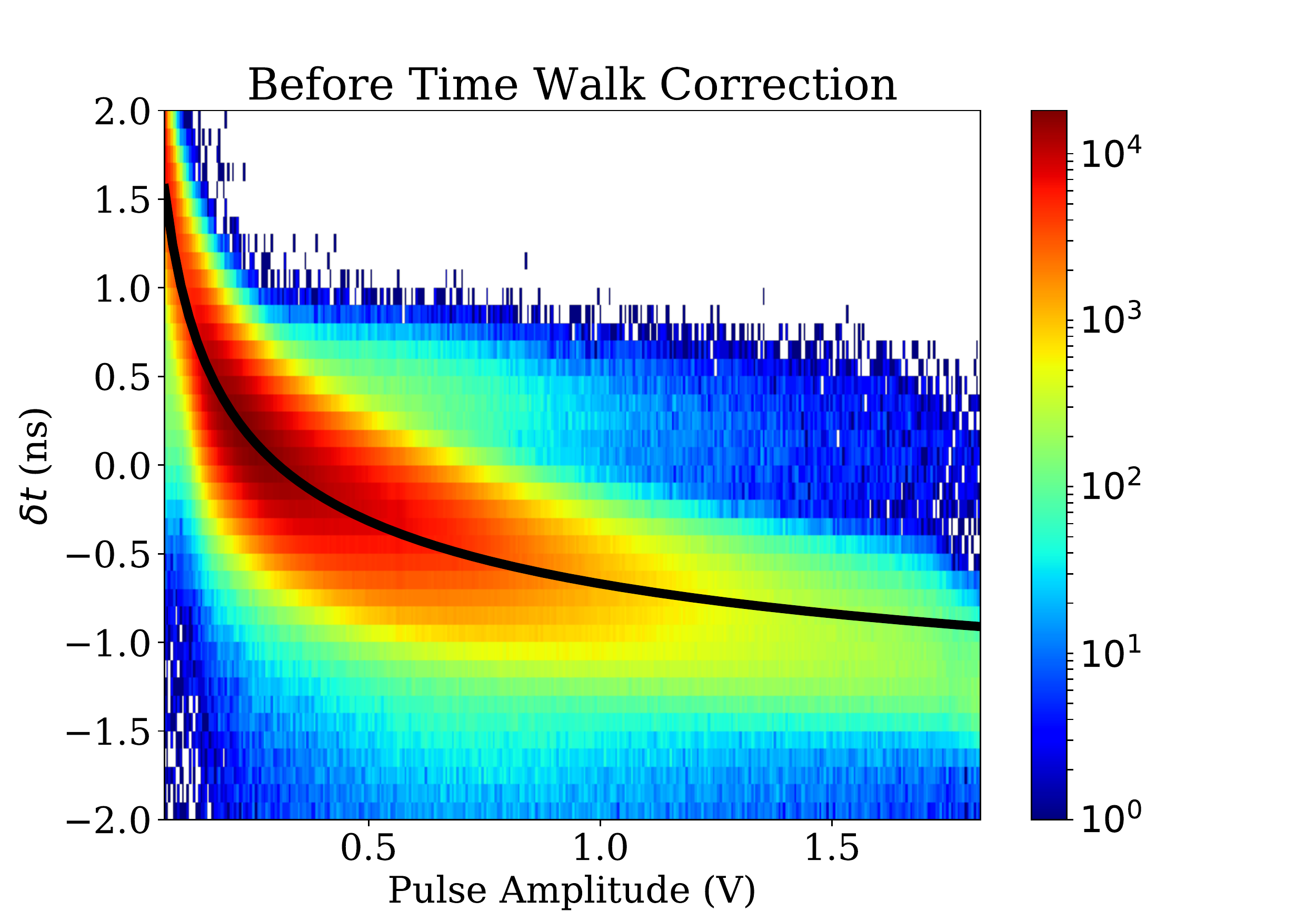}}\hfill
		\subfigure{\includegraphics[width=0.5\textwidth]{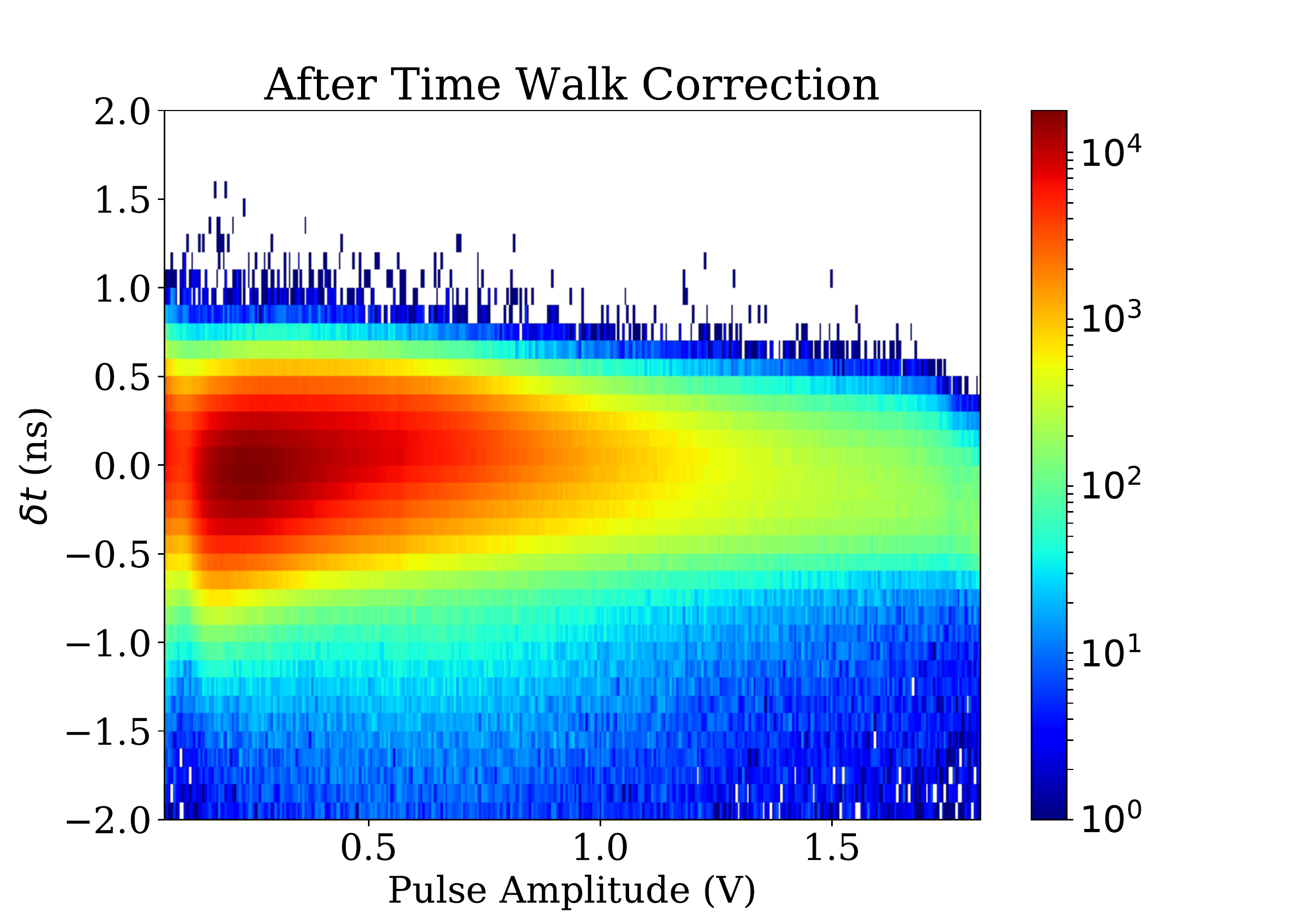}}
		\caption{Left: Single paddle time-walk spectrum; the line shown is the 
		fitted function used to determine the correction factors. Right: after 
		time-walk correction.  Plotted on the vertical axis is $\delta t$ and 
		on the horizontal axis is the corresponding pedestal subtracted pulse 
		amplitude spectrum.} 
		\label{fig:time_walk} 
	\end{figure*}
This correlation can be described empirically by the function given by 
Eq.~\ref{eq:tw_corr_func_form}~\cite{esmith_bcal} where $a$ and 
$a^{\rm thresh}_{i}$ are the pulse amplitude and discriminator threshold 
respectively, and $c0_{i}, c1_{i}, c2_{i}$ are the fit parameters. 
	\begin{equation} \label{eq:tw_corr_func_form}
		f^{w}_{i}\left(a/a^{\rm thresh}_{i}\right) = c0_{i} + 
		\frac{c1_{i}}{(a/a^{\rm thresh}_{i})^{c2_{i}}}
	\end{equation} 

The most probable value (MPV) of the pulse amplitude spectra was chosen as the 
reference point where the time-walk correction is defined to be zero.  
Fig.~\ref{fig:time_walk} (right) illustrates the effect on the time 
difference spectrum ($\delta t$) as a result of the applied time-walk 
corrections. 

\subsection{Propagation Time Corrections} \label{sec:calib_ptc}

The time between the production of scintillation light in a ST scintillator 
paddle and detection by the SiPM depends on the hit location along the paddle 
and is discussed below. 

The EJ-200 scintillator material has a refractive index of 1.58 
\cite{ej200_specs} and the corresponding speed of light in that medium is 
$\mathrm{19\ cm/ns}$.  The observed effective velocity is slower.  Correcting 
for this light propagation in the scintillator is necessary since the ST 
paddles are 60~cm long.  Studies showed that the effective velocity of light 
depends on the region along the paddle where the hit occurred.  The propagation 
time corrections were conducted with well-defined reconstructed charge particle 
tracks.  Further details regarding the event and track selection are found in 
Ref.~\cite{pooser16}. 

The propagation time $T^{\rm ST}_{\rm prop}$ is determined by  
Eq.~\ref{eq:st_prop_time} where $T^{\rm ST}_{\rm hit}$ is the time-walk 
corrected hit time, $T^{\rm ST}_{\rm flight}$ is the flight time from the track 
vertex to the ST intersection point, and $T^{\rm BB}_{\rm vertex}$ is the track
vertex time.
	\begin{equation} \label{eq:st_prop_time}
		T^{\rm ST}_{\rm prop} = T^{\rm ST}_{\rm hit} - T^{\rm ST}_{\rm flight} 
		- T^{\rm BB}_{\rm vertex}
	\end{equation}
The $z$-coordinate of the track's intersection point with the ST $(z^{\rm 
ST}_{\rm hit})$ are determined by the detector geometry as well as the distance 
$d^{\rm ST}_{\rm hit}$ of this intersection point and the SiPM.

The propagation times were determined in three distinct regions corresponding 
to the three geometrical sections of the ST: the straight, bend, and nose  
regions.  The propagation times in these regions were fit with a linear  
function given by Eq. \ref{eq:pt_func_form} where $j$ indicates which region in 
the $i^{th}$ paddle is being fit and $A$ and $B$ are fit parameters. 
	\begin{equation} \label{eq:pt_func_form}
		f^{i}_{j}(z) = A^{i}_{j} + B^{i}_{j} \cdot z
	\end{equation} 
Figure~\ref{fig:proptimeuncorr} (left) illustrates the correlation between the 
propagation time and the distance from the SiPM with $T^{\rm ST}_{\rm prop} = 
0.0\ 
\mathrm{ns}$ when $d^{\rm ST}_{\rm hit} = 0.0\ \mathrm{cm}$. 
	\begin{figure*}[!htb]
		\centering
		\subfigure{\includegraphics[width=0.5\textwidth]{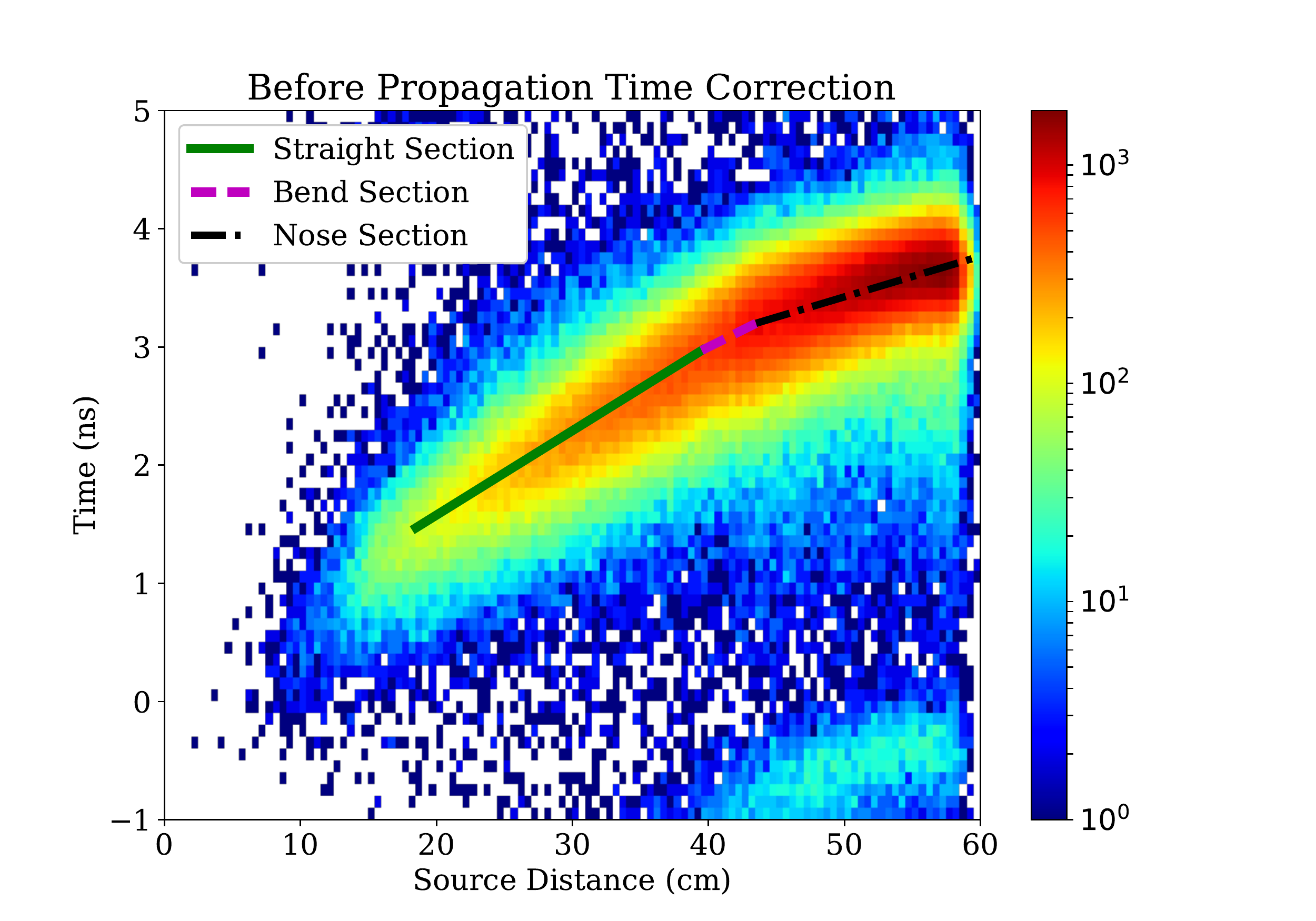}}\hfill
		\subfigure{\includegraphics[width=0.5\textwidth]{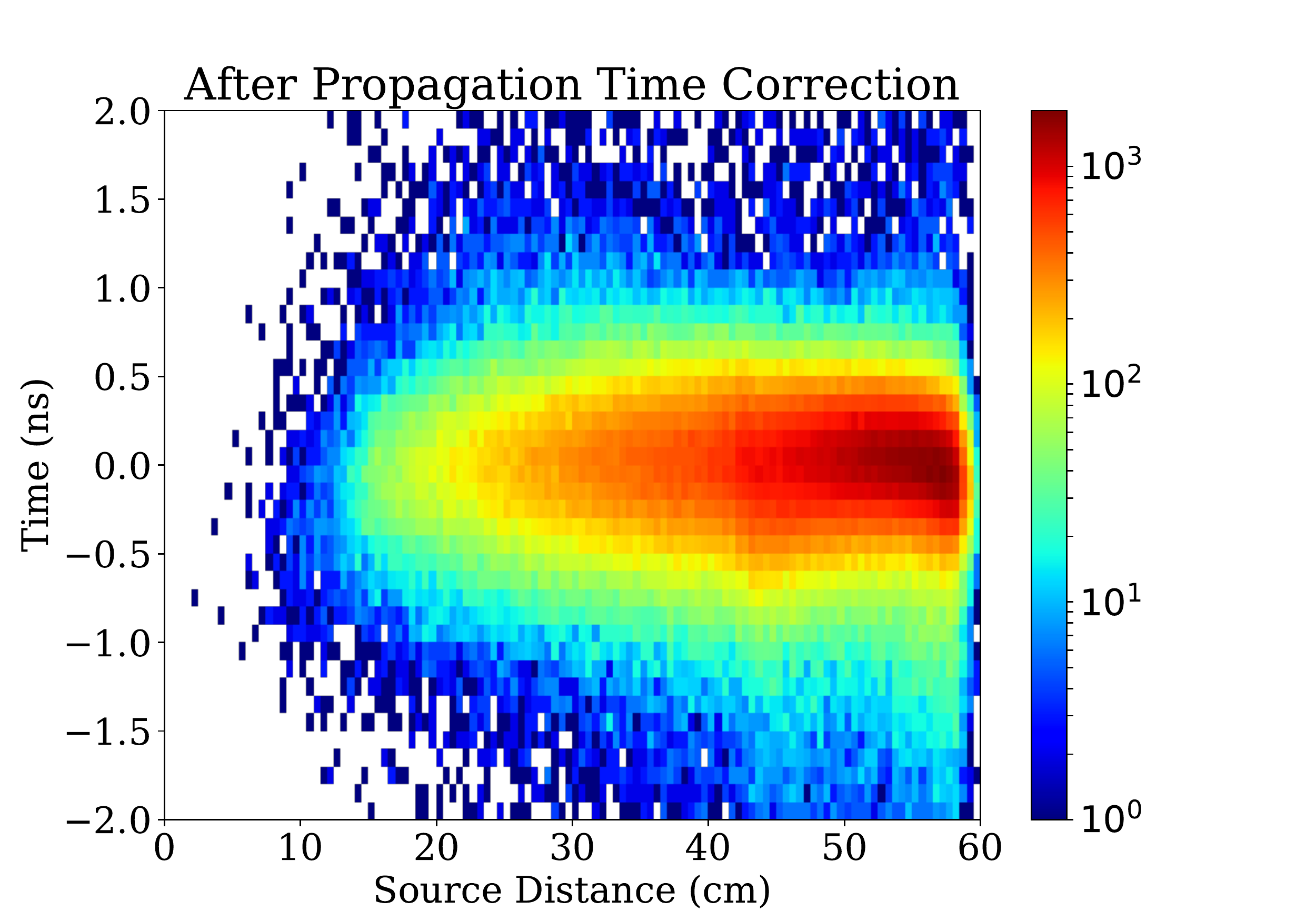}}
		\caption{Left: Single paddle propagation time correlation.  
		$T^{\rm ST}_{\rm prop}$ is plotted on the vertical axis and $d^{\rm 
		ST}_{\rm hit}$ is 
		plotted along the horizontal axis. There is a clear correlation between 
		the time when optical photons are detected by the SiPM and the location 
		of the scintillation light along the length of the paddle. Right: 
		Single paddle propagation time after correction.}
		\label{fig:proptimeuncorr}
	\end{figure*} 
Figure ~\ref{fig:proptimeuncorr} (right) illustrates the corrected time.

\subsection{Attenuation Corrections} \label{sec:calib_ac}

To measure attenuation in the scintillators, charged tracks were selected in a 
manner similar to that discussed in Sec.~\ref{sec:calib_ptc}.  The uncorrected 
energy deposition $(dE_{M})$ per unit length $(dx)$ versus the track momentum 
$(p)$ for tracks intersecting to the ST are shown in Fig.~\ref{fig:dEdx_vs_p}.  
It is clear that no reliable PID can occur for tracks with $p > 0.6\ 
\mathrm{GeV/c}$ without further corrections. 

The pulse integral (PI) data, normalized to the path length $dx$ of the track 
in the scintillator medium, were binned in 3.5~cm $z^{\rm ST}_{\rm hit}$ bins 
along the length of the paddle. The MPV of the PI was extracted utilizing an 
empirical function given by Eq.~\ref{eq:mpv} where $p_{0},\ p_{1},\ p_{2}$ are 
the fit parameters. 
	\begin{equation}\label{eq:mpv} 
		f(z)  = p_0 e^{(-p_1(z - p_2))} \times (1+ \tanh(p_1(z - p_2))) 
	\end{equation}
A fit to the data in a single 3.5~cm $z^{\rm ST}_{\rm hit}$ bin is illustrated 
in Fig.~\ref{fig:mpv_fit}. 
	\begin{figure}[!htb]
		\centering
		\includegraphics[width=1.0\columnwidth]{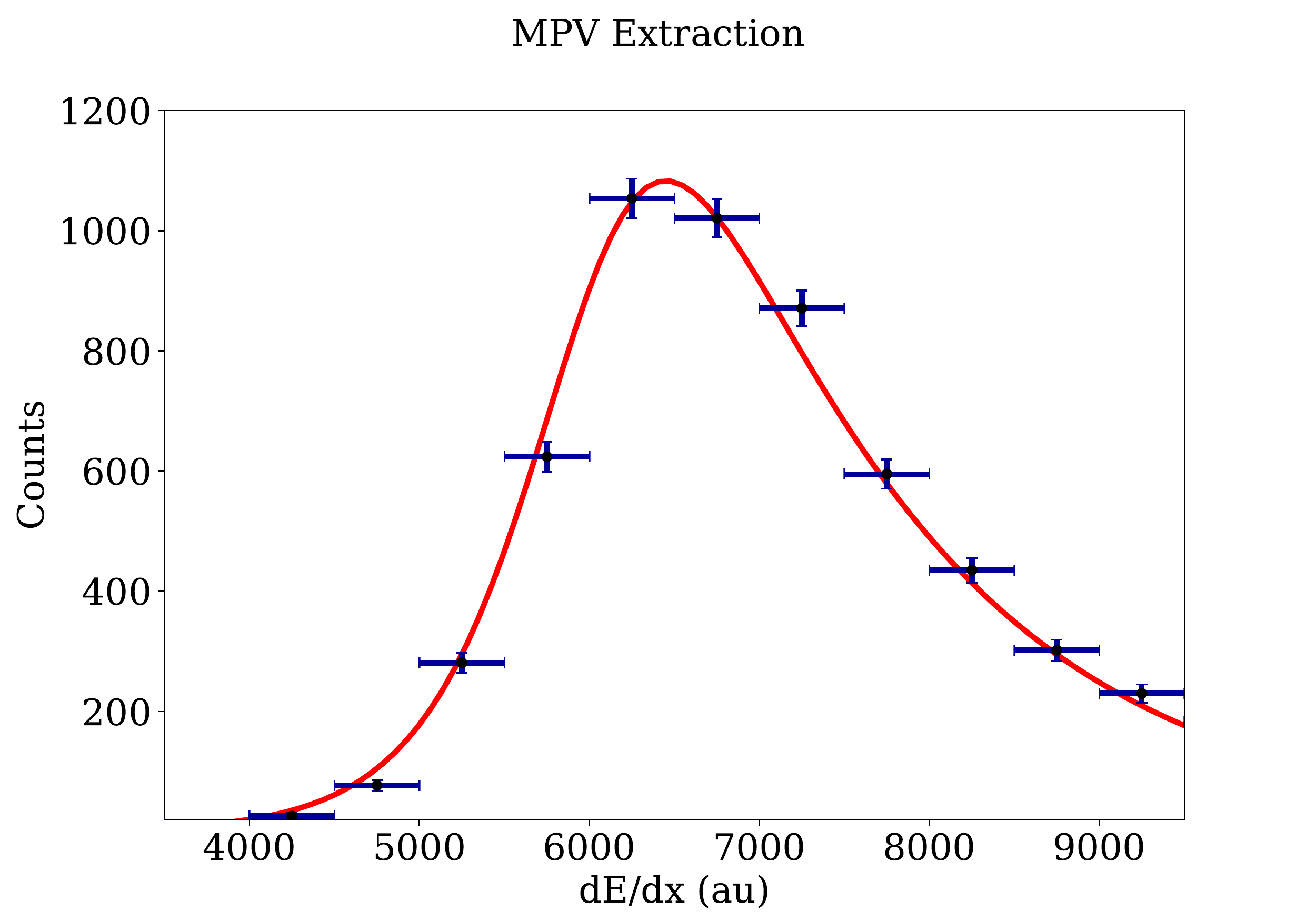}
		\caption{Pulse integral integral data normalized to the the track 
		length in the scintillator medium for a single 3.5~cm bin along the 
		paddle length.} 
		\label{fig:mpv_fit}
	\end{figure}
The MPV was extracted analytically and then plotted against the average value 
for each $z^{\rm ST}_{\rm hit}$ bin as shown in Fig.~\ref{fig:attfits}. 
	\begin{figure}[!htb]
		\centering
		\includegraphics[width=1.0\columnwidth]{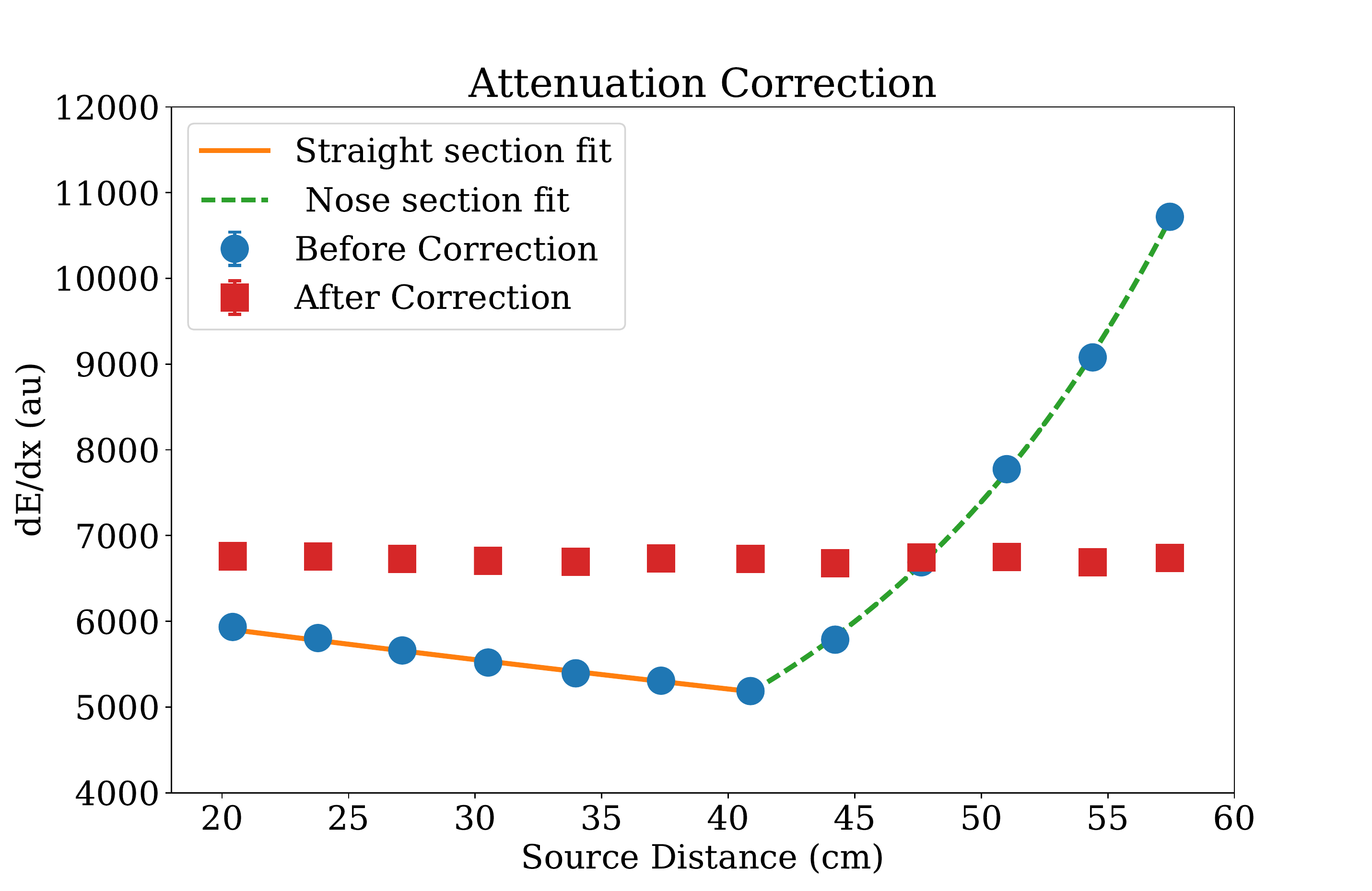}
		\caption{Fits to the attenuation data.}
		\label{fig:attfits}
	\end{figure}

In order to characterize the photon attenuation, the straight and nose regions 
were treated independently.  The piecewise continuous function given by 
Eq.~\ref{eq:attn_fit_disc_pw} was selected to fit the data where the  
intersection $Z^{i}_{b}$ (or correction boundary) of the two exponential fit  
functions was fixed and is shown in Fig.~\ref{fig:attfits}.
	\begin{equation} \label{eq:attn_fit_disc_pw}
		f_{c}^{i}(z) = 
		\begin{cases} 
		A^{i}_{S}e^{-B^{i}_{S} \cdot z} & z \leq Z^{i}_{b}\ \mathrm{cm} \\
		A^{i}_{N}e^{B^{i}_{N} \cdot z} + C^{i}_{N} & z > Z^{i}_{b}\ \mathrm{cm} 
		\\
		\end{cases}
	\end{equation}
In Eq.~\ref{eq:attn_fit_disc_pw}, the subscripts $S$ and $N$ denote the 
straight and nose sections respectively while $A^{i}, B^{i},$ and $C^{i}$ are 
the fit parameters for the $i^{th}$ paddle.

An attenuation correction factor $R^{i}(z)$ is applied to the deposited energy 
measurement per unit track-length $(dE_{M} / dx)$ to give the corrected energy 
deposition per unit track length $(dE^{i}_{C}(z) / dx)$ for paddle $i$ and is 
given by Eq.~\ref{eq:de_corr_init} where the subscripts $C$ and $M$ are the 
corrected and measured quantities respectively.  
	\begin{equation} \label{eq:de_corr_init}
		\dfrac{dE^{i}_{C} (z)}{dx} = \dfrac{dE_{M}}{dx} \cdot R^{i}(z) = 
		\dfrac{dE_{M}}{dx} \cdot \dfrac{f^{i}_{c}(0)}{f^{i}_{c}(z)}
	\end{equation}
After attenuation corrections are applied, particle separation is greatly 
improved. This will be discussed further in Sec.~\ref{sec:perform}.
\section{Performance} \label{sec:perform}

The increase in light output as a function of hit position along the ST 
detector during nominal \gx{} beam conditions is illustrated in 
Fig.~\ref{fig:pippvszint}.
	\begin{figure}[!htb]
		\centering
		\includegraphics[width=1.0\columnwidth]{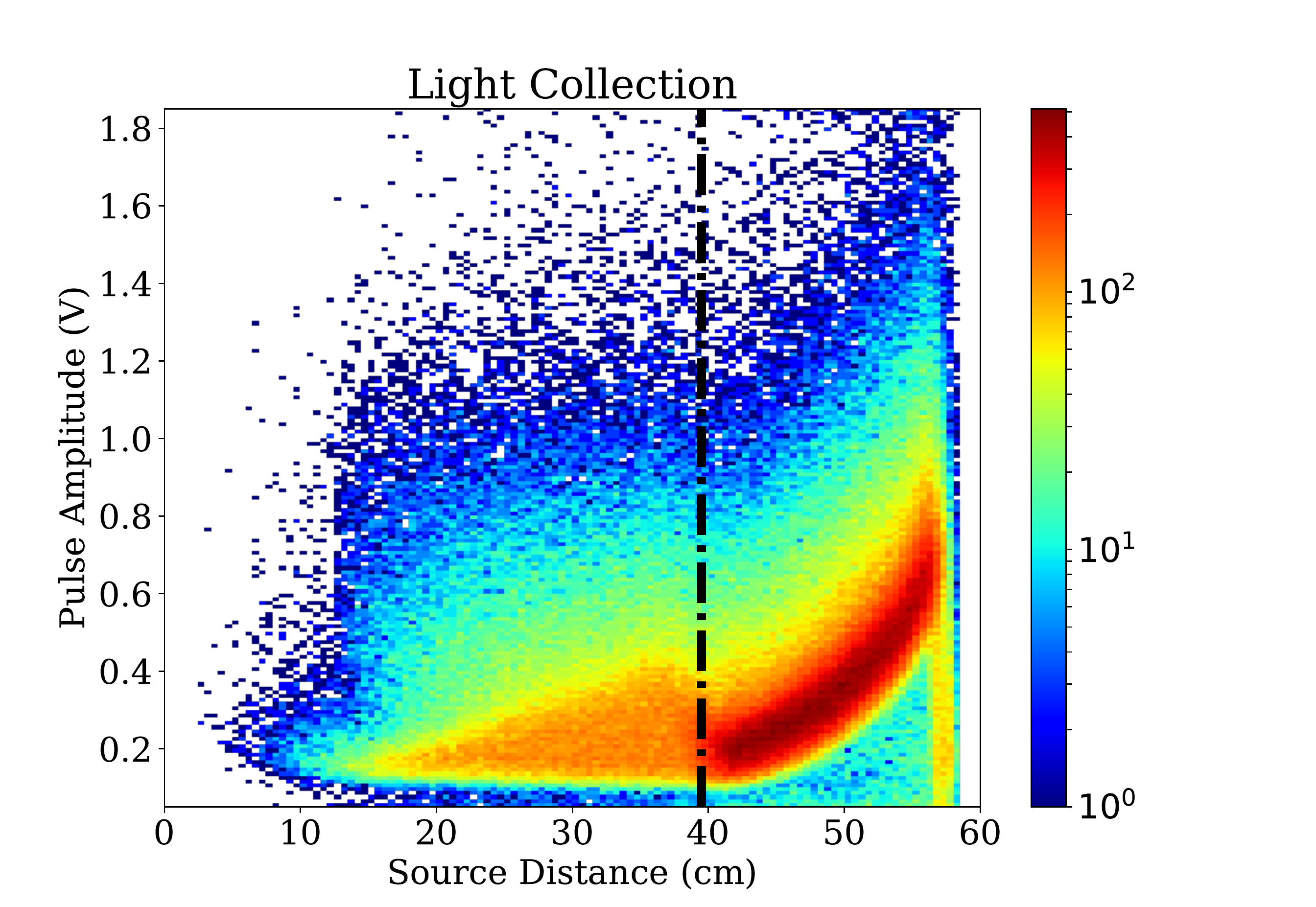}
		\caption{Typical FADC250 pulse amplitude spectrum versus the 
			$z$-component of charged tracks intersecting the ST for an 
			individual ST sector. The vertical line indicates the start of the 
			tapered nose section.}
		\label{fig:pippvszint} 
	\end{figure}
This is advantageous because the majority of the charged tracks produced  
intersect the ST in the forward region.

With the attenuation corrections discussed in Sec.~\ref{sec:calib_ac} applied 
to the data, the PID capabilities of the ST were improved.  Figure 
~\ref{fig:dEdx_vs_p} illustrates the PID capability of charged tracks  
intersecting the ST.
	\begin{figure*}[!htb]
		\centering
		\subfigure{\includegraphics[width=0.5\textwidth]{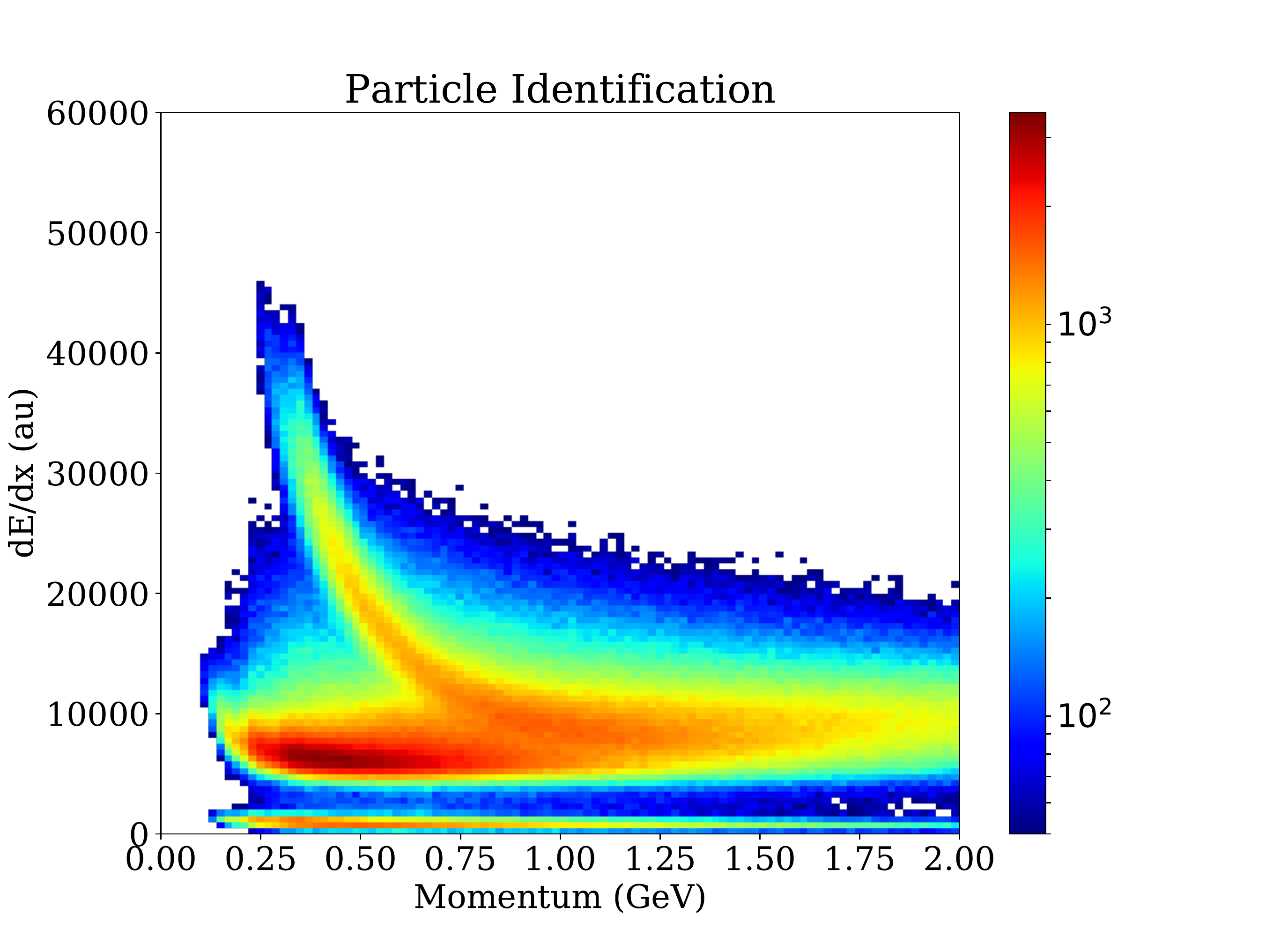}}\hfill
		\subfigure{\includegraphics[width=0.5\textwidth]{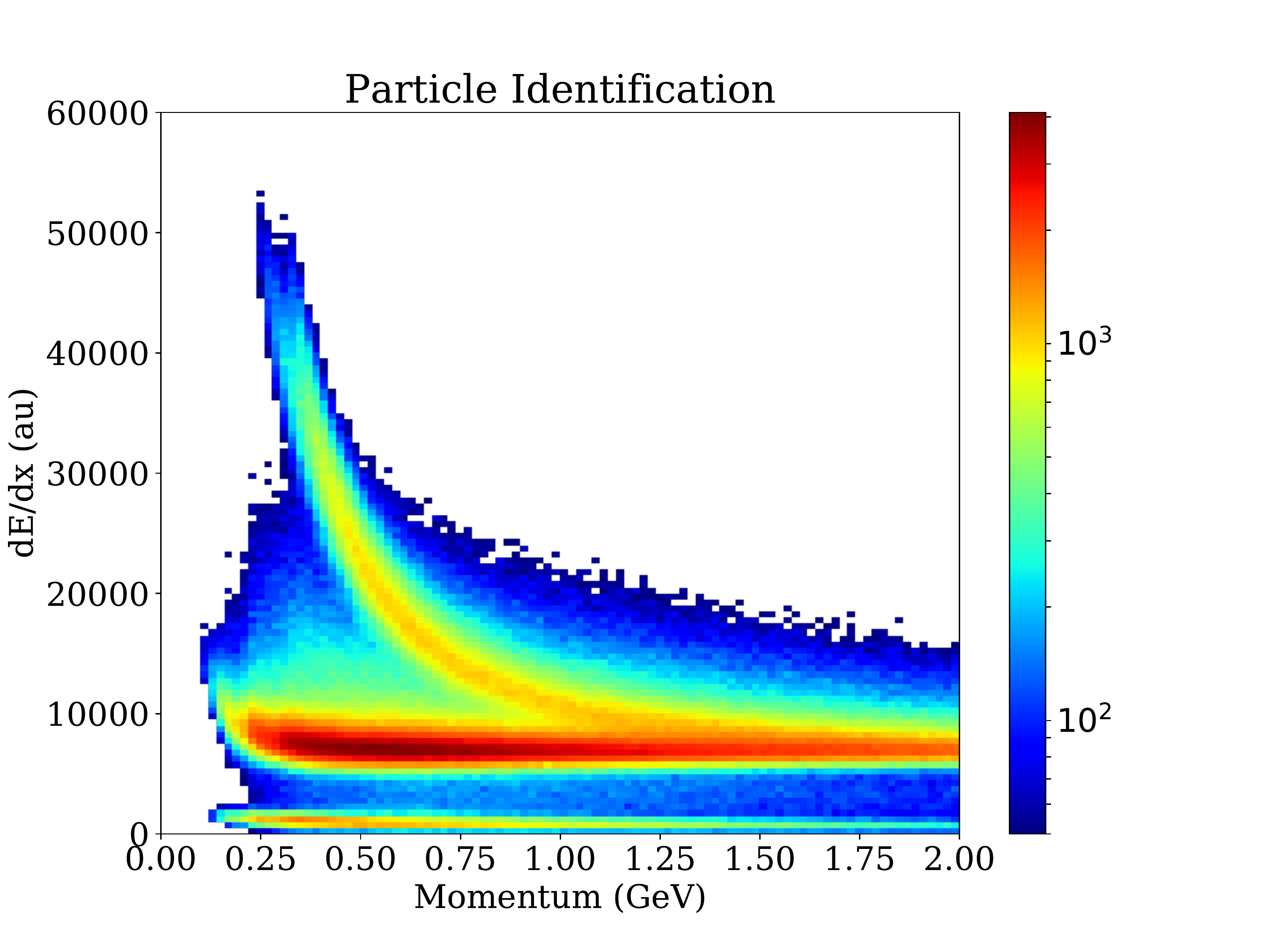}}
		\caption{Left: Typical uncorrected $dE/dx\ vs.\ p$ distribution. Right: 
		Corrected $dE/dx\ vs.\ p$ distribution.  The ``banana band'' 
		corresponds to protons while the horizontal band corresponds to 
		electrons, pions, and kaons. It is clear that after the corrections 
		have been applied, pion/proton separation is achievable for tracks with 
		$p < 0.9\ \mathrm{GeV/c}$.} 
		\label{fig:dEdx_vs_p}
	\end{figure*}  
Protons can be separated from other hadrons with momenta up to 0.9 GeV/c which 
is a factor 1.5 improvement relative to the uncalibrated data.  The PID 
capabilities of the ST extend the identification of low momentum protons that 
do propagate through the central drift chambers.

The ST was used to determine the time of the interaction of a beam photon with 
the $\mathrm{LH_{2}}$ target after the time-walk and propagation time 
corrections discussed in Sec.~\ref{sec:calib_tw} and \ref{sec:calib_ptc} were 
applied. The interaction time can be determined independently from a timing 
signal originating from the accelerator’s RF system, the latter with very high 
precision. The time difference between the ST time and the machine RF time is 
shown in Fig.~\ref{fig:st_time_res}. The accelerator can be run in mode where 
the time separation between beam bunches is 2~ns, a separation indicated in the 
figure. One application of the ST is to distinguish particles from different RF 
buckets on the basis of timing.
	\begin{figure}[!htb]
		\centering
		\includegraphics[width=1.0\linewidth]{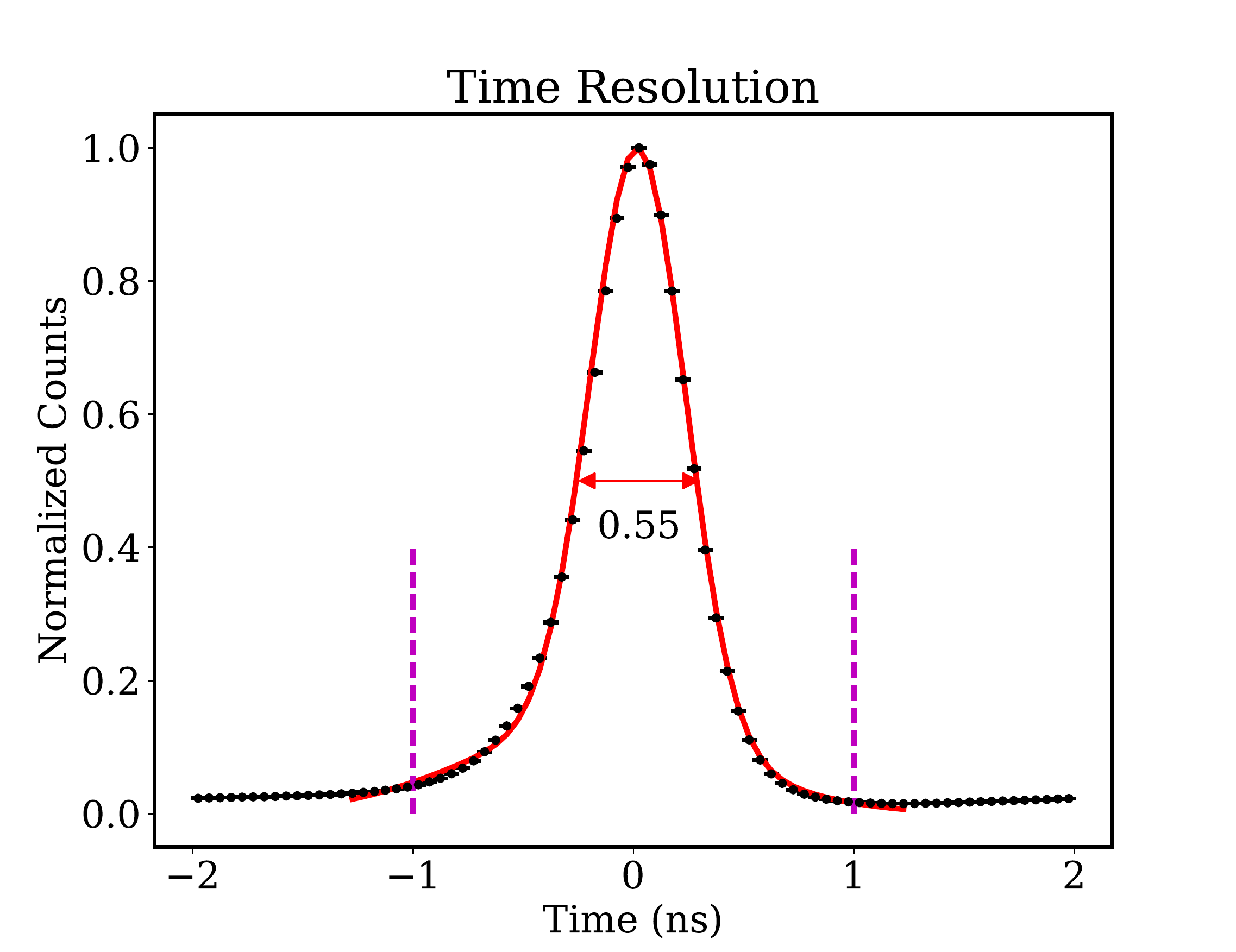}
		\caption{Time resolution for one paddle with its full width half 
		maximum value indicated in ns.  The x-axis is the time difference 
		between $T^{\rm ST}_{\rm vertex}$ and $T^{\rm BB}_{\rm vertex}$. The 
		vertical lines indicate the cuts used to identify a 500~MHz beam bunch.}
		\label{fig:st_time_res}
	\end{figure}  

The measured distribution is fit with the sum of two Gaussians and the full 
width at half maximum (FWHM) of the resulting curve is calculated. Also, the 
fraction of the area of the curve within $\pm$1~ns of zero is calculated. Fits 
were carried out for the three individual geometrical regions and for all 30 
paddles. The results are shown in Table~\ref{tab:time_res_section}.
	\begin{table}[htbp]
		\centering
		\begin{tabular}{@{} l *4c @{}}
			\toprule
			\multicolumn{1}{c}{\textbf{Section}}    & \textbf{All}  & \textbf{Straight}  & \textbf{Bend}  & \textbf{Nose}  \\ 
			\midrule
			$\mathbf{FWHM}$ & 550~ps & 690~ps & 700~ps & 450~ps \\ 
			\textbf{Fraction} & 93\% & 92\% & 91\% & 94\% \\\bottomrule 
		\end{tabular}
		\caption{Average time resolutions (FWHM) and event fractions within a $\pm$ 1~ns window for all 30 ST sectors by independent geometrical regions.}
		\label{tab:time_res_section}
	\end{table}  

The ST exhibited uniformity in time resolution among all sectors of the ST. The 
high overall event fraction and good time resolution in the data for all 
sections combined is due to the majority of events intersecting the ST in the 
nose section.  It is clear from Table~\ref{tab:time_res_section} that 
measurements made with beam data also exhibit improved light collection, and 
thus improved time resolution, in the nose region.

Approximately four years of operation have elapsed since the paddles were first 
tested on the bench at Florida International University (FIU).  Prior 
experience with scintillators indicates that degradation in time resolution as 
a result of mishandling will be visible in a matter of weeks.  No degradation 
in time resolution has been observed and the ST is still performing well below 
design resolution.

\section{Conclusion} \label{sec:conclusion}

The \gx{} Start Counter was designed and constructed at Florida International 
University for use in Hall D at TJNAF. It provides separation of the 500 MHz 
photon beam bunch structure delivered by the CEBAF to within 94\% accuracy.  It 
is the first ``start counter'' detector to utilize magnetic field insensitive 
SiPMs as the readout system.  Despite the many design and manufacturing 
complications, the ST has proven to have performed well beyond the design 
resolution of 825~ps (FWHM) with an average measured resolution of 550~ps 
(FWHM).  Furthermore, the capabilities of the ST make it a viable candidate to 
assist in particle identification. 

The unique geometry of the ST nose section has illustrated the advantage of 
tapering trapezoidal geometry in thin scintillators.  Through simulation, tests 
on the bench, and analysis of data obtained with beam, it has been definitively 
demonstrated that this geometry results in a phenomenon in which the amount of 
light detected increases as the scintillation source moves further downstream 
from the readout detector. 

Since its installation in Hall D during the Fall 2014 commissioning run, the ST 
has shown no measurable signs of deterioration in performance.  This suggests 
that the ST scintillators are void of crazing and will most likely be able to 
meet and exceed the design performance well beyond the scheduled run periods 
associated with the \gx{} experiment. 

It is planned to incorporate the ST into the level 1 trigger of the \gx{} 
experiment for high luminosity running when there will be $5 \times 10^{7}\ 
\gamma/s$ in the coherent peak.  Preliminary studies suggest that while 
operating at rates in excess of 300 kHz per paddle, the ST exhibits a high 
efficiency $(> 95\%)$.  Thus, in combination with the calorimeters the ST has 
the potential to provide good suppression of electromagnetic background 
via incorporation into the level 1 trigger of the experiment.  Furthermore, the 
ST's high degree of segmentation has demonstrated suppression of various 
background contributions associated with complex topologies while 
simultaneously providing precision timing information for reconstructed charged 
particles in \gx{}. 
\section{Acknowledgments}

The authors would like to graciously thank the plethora of Jefferson Lab staff 
members in both the Engineering division and Hall-D.  Their numerous and 
invaluable contributions allowed for the Start Counter project to come to 
fruition.  The authors would also like to extend their gratitude to the entire 
\gx{} Collaboration who provided fruitful ideas and advice throughout the many 
stages of the project.  Work at Florida International University was supported 
in part by the Department of Energy under contracts DE-FG02-99ER41065 and 
DE-­SC00-13620. Furthermore, this material is based upon work supported by the 
U.S. Department of Energy, Office of Science, Office of Nuclear Physics under 
Contract No. DE-AC05-06OR23177.  
\endgroup

\newpage
\section*{References}
\bibliography{bibliography/bibliography}

\end{document}